\documentclass[12pt]{article}
\usepackage{a4}
\usepackage{latexsym} 
\usepackage{amssymb}  
\usepackage{graphicx}
\usepackage{epsfig}

\newcounter{zyxabstract}     
\newcounter{zyxrefers}        

\newcommand{\newabstract}
{\newpage\stepcounter{zyxabstract}\setcounter{equation}{0}
\setcounter{footnote}{0}}

\newcommand{\rlabel}[1]{\label{zyx\arabic{zyxabstract}#1}}
\newcommand{\rref}[1]{\ref{zyx\arabic{zyxabstract}#1}}

\renewenvironment{thebibliography}[1] 
{\section*{References}\setcounter{zyxrefers}{0}
\begin{list}
{[\arabic{zyxrefers}]}
{\usecounter{zyxrefers}\setlength{\parindent}{0cm}\setlength{\itemsep}{0cm}}

}
{\end{list}}
\newenvironment{thebibliographynotitle}[1] 
{\section*{References}\setcounter{zyxrefers}{0}
\begin{list}{[\arabic{zyxrefers}]}
{\usecounter{zyxrefers}\setlength{\parindent}{0cm}\setlength{\itemsep}{-1.5mm}}}
{\end{list}}

\renewcommand{\bibitem}[1]{\item\rlabel{y#1}}
\renewcommand{\cite}[1]{[\rref{y#1}]}      
\newcommand{\citetwo}[2]{[\rref{y#1},\rref{y#2}]}
\newcommand{\citethree}[3]{[\rref{y#1},\rref{y#2},\rref{y#3}]}
\newcommand{\citefour}[4]{[\rref{y#1},\rref{y#2},\rref{y#3},\rref{y#4}]}

\newcommand{\citesix}[6]
{[\rref{y#1},\rref{y#2},\rref{y#3},\rref{y#4},\rref{y#5},\rref{y#6}]}

\newcommand{\vs}{}
\begin{document}
\begin{titlepage}
\begin{flushright} 
{\small
HISKP-TH-05/03\\
FZJ-IKP-TH-2005-03\\
LU-TP 05-04
}
\end{flushright}

\begin{center}
{\huge\bf EFFECTIVE FIELD THEORIES\\[2.1mm]IN NUCLEAR, PARTICLE\\[2.1mm]
 AND ATOMIC PHYSICS$^{*}$}
\\[1cm]
337. WE-Heraeus-Seminar\\
Physikzentrum Bad Honnef, Bad Honnef, Germany\\
December 13 --- 17, 2004\\[1cm]
{\bf Johan Bijnens}$^{1}$,  {\bf Ulf-G. Mei\ss ner}$^{2,3}$ 
and {\bf Andreas Wirzba}$^3$\\[0.3cm]
$^1$Department of Theoretical Physics, Lund University\\
S\"olvegatan 14A, S22362 Lund, Sweden\\[0.3cm]
$^2${Universit\"at Bonn , Helmholtz-Institut f\"ur Strahlen- und Kernphysik (Theorie)\\ 
D-53115 Bonn, Germany}\\[0.3cm]
$^3${Forschungszentrum J\"ulich, Institut f\"ur Kernphysik (Theorie)\\ 
D-52425 J\"ulich, Germany}\\[1cm]
{\large ABSTRACT}
\end{center}
These are the proceedings of the workshop on ``Effective Field Theories in Nuclear, Particle and Atomic Physics''
held at the Physikzentrum Bad Honnef of the
Deutsche Physikalische Gesellschaft, Bad Honnef, Germany from 
December 13 to 17, 2005. The workshop concentrated on
Effective Field Theory in many contexts. A first part was
concerned with Chiral Perturbation
Theory in its various settings and explored strongly
its use in relation with lattice QCD.
The second part consisted of progress in effective field theories
in systems with one, two or more nucleons as well as atomic physics.
Included are a short contribution per talk.

\noindent\rule{6cm}{0.3pt}\\
\footnotesize{$^*$ This workshop was funded by the WE-Heraeus foundation.
This
research is part of the EU Integrated Infrastructure Initiative Hadron
Physics Project under contract number RII3-CT-2004-506078.
Work supported in part by DFG (SFB/TR 16 ``Subnuclear Structure of Matter''),
the European Union TMR network,
Contract No. 
HPRN-CT-2002-00311  (EURIDICE),
Forschungszentrum J\"ulich and the Swedish Research Council.}
\end{titlepage}

\section{Introduction}
The use of effective field theory techniques is an ever growing approach
in various fields of  theoretical physics. 
Along with the continuing application of 
Chiral Perturbation Theory and Nuclear Effective Field Theory, more
recently, chiral symmetry in lattice QCD (chiral fermions, chiral
extrapolations, finite volume effects) and atomic few-body systems have been 
the focus of many
investigations.
We therefore decided to organize the next topical workshop, with an emphasis 
of bringing together people from these various communities. 
This meeting followed the series of workshops in Ringberg (Germany), 1988,
Dobog\'ok\"o (Hungary), 1991, Karreb\ae ksminde (Denmark), 1993,
Trento (Italy), 1996 and Bad Honnef (Germany), 1998 and 2001.
All these workshops shared the same features,
about 50 participants, a fairly large amount of time devoted to discussions
rather than presentations and an intimate environment with lots of
discussion opportunities.

This meeting took place in late fall 2004 in the Physikzentrum Bad Honnef
in Bad Honnef, Germany and the funding provided by the 
WE--Heraeus--Stiftung allowed us to provide for the local
expenses for all participants and to support the travel of a fair amount of
participants. The WE-Heraeus foundation also provided
the administrative support for the workshop in the person of the
able secretary Heike Uebel. We extend our sincere gratitude to the WE-Heraeus
Stiftung for this support. We would also like to thank the staff of the
Physikzentrum for the excellent service given to us during the workshop
and last but not least the participants for making this an exciting
and lively meeting.

The meeting had 58 participants whose names, institutes and email addresses
are listed below. 48 of them presented results in presentations of various
lengths.
A short description of their contents and a list of the most relevant
references can be found below. As in the previous three of these workshops we
felt that this was more appropriate a framework than full-fledged proceedings.
Most results are or will soon be published and available on the archives,
so this way we can achieve speedy publication and avoid duplication of
results in the archives.

Below follows first the program, then the list of participants 
followed by the abstracts of the talks. Most of them can also be
obtained from the workshop website 

\centerline{{\tt http://www.itkp.uni-bonn.de/$\sim$eft04}~.}

\vspace{1.5cm}

\hfill Johan Bijnens, Ulf-G. Mei\ss ner and Andreas Wirzba

\newpage

\section{Program}
\begin{tabbing}
xx:xx \= A very very very long name \= \kill
{\bf Monday, December 13th 2004}\\[3mm]
{\it  Early Afternoon  Session}\\
Chair: Johan   Bijnens    \>       \> {\bf  Chiral Perturbation Theory} \\

  14:00 \>    Ulf-G. Mei{\ss}ner /    \>  Introductory Remarks \\
        \>    Ernst Dreisigacker  \> \\
14:20 \>        Aneesh Manohar (La Jolla) \> 
   $1/N_c$ and pentaquarks\\
 15:10 \>
        Heiri Leutwyler (Bern)\>
        How well do we understand the interaction among\\
  \>\> the pions at low energies? \\
15:50 \>\>{\em  Coffee}\\
{\it   Late Afternoon Session}\\
Chair: Andreas Wirzba \>\> {\bf Chiral Perturbation Theory}\\
16:20 \>
        Uwe-Jens Wiese (Bern)\>         Can one see the number of colors ?\\
17:00 \>
        S\a'ebastien Descotes-Genon  \>
       Sea quark effects in three-flavour chiral\\
 \>(Orsay)\> perturbation theory\\
       
17:40 \>
        Barry Holstein \>       Linearized GDH sum rule\\
 \>(Massachusetts)\>\\
18:20\>\>{\em End of Session}\\
18:30\>{\em Dinner --- Invitation by the Wilhelm und Else Heraeus-Stiftung }
\\[3mm]
{\bf  Tuesday, December 14th, 2004}\\[3mm]
{\it  Early Morning Session}\\
        Chair: Ulf-G. Mei\ss ner \>\>{\bf  Chiral Perturbation Theory}\\
09:00 \>
        Akaki Rusetsky (Bonn)\>   Determination of the $\pi N$ scattering
        lengths from\\ \>\> the experiments on pionic deuterium\\
09:40\>
        Udit Raha (Bonn)\>
        Spectrum and decay constants of kaonic hydrogen\\
10:00 \>
        Robin Ni\ss ler (Bonn) \>    Anomalous decays of $\eta$ 
      and $\eta'$ with coupled channels \\
10:20 \>\> {\em Coffee}\\
{\em Late Morning Session}\\
        Chair: Ulf-G. Mei\ss ner\>\>{\bf   Chiral Perturbation Theory}\\
10:50\>
        Bastian Kubis (Bonn)  \>  Radiative Ke3 decays revisited \\
11:15\>
        Bashir Moussallam (Orsay)  \>   Electromagnetic LEC's and QCD n-point\\
 \>\> functions resonance models \\
11:40\>
        Jos\a'e A. Oller (Murcia)\>  Scalar mesons in D hadronic decays \\
12:15\>
        Eulogio Oset (Valencia)    \>   Chiral dynamics of baryon resonances\\
12:55\>\>{\em End of Session}\\
13:00\>\>{\em Lunch}\\

{\em Early Afternoon Session}\\
        Chair: Johan Bijnens \>\>{\bf   Chiral Perturbation Theory}\\
14:20\>
        Marc Knecht (Marseille)    \>     The Dalitz decay $\pi^0\to e^+ e^- \gamma$ \\
15:00\>
        Gilberto Colangelo (Bern)  \>     The pion vector form factor and the muon\\ \>\> anomalous magnetic moment \\
15:40\>
        Timo L\"ahde (Lund)   \>    Partially quenched chiral perturbation theory\\ \>\> at NNLO \\
16:05\>\>{\em Coffee}\\
{\em Late Afternoon Session}\\
        Chair: Johan Bijnens \>\>{\bf   Chiral Perturbation Theory}\\
16:35\>
        Toni Pich (Valencia)  \>   Effective lagrangians in the resonance region \\
17:15\>
        Peter Bruns (Bonn)   \>    Infrared regularization for spin-1 fields\\
17:40\>
        Joaquim Prades (Granada)  \>       A large Nc hadronic model\\
18:05\>
        Matthias Frink (Bonn) \>   Baryon masses in cut-off regularized ChPT\\
18:25\>\>        
{\em End of Session}\\
18:30\>\>        
{\em Dinner}\\[3mm]
{\bf Wednesday, December 15th, 2004}\\[3mm]
{\em Early Morning Session}\\
        Chair: Andreas Wirzba  \> \> {\bf Nuclear Effective Theory}\\
09:00\>
        Evgeny Epelbaum  \>  Chiral dynamics in few-nucleon systems\\
\>(Newport News)\\
09:40\>
        Bira van Kolck (Tucson)   \>       Charge-symmetry-breaking nuclear forces \\
10:15\>
        Jambul Gegelia (Mainz)  \> Consistency of Weinberg's approach to the \\
\>\>few-nucleon problem in EFT \\
10:35\>\>        
{\em Coffee}\\
{\em Late Morning Session}\\
        Chair: Andreas Wirzba  \>\>{\bf  Nuclear Effective Theory}\\
11:05\>
        Andreas Nogga (J\"ulich) \>  Renormalization of the 1pi exchange interaction\\ \>\> in higher partial waves \\
11:30\>
        Enrique Ruiz Arriola \>  Renormalization group approach to NN-scattering\\ \>  (Granada)\> with pion exchanges: removing the cut-offs\\
11:55\>
        Luca Girlanda (Trento) \>  Chiral perturbation theory for heavy nuclei\\
12:20\>
        Norbert Kaiser (Garching)   \>     Chiral dynamics of nuclear matter: Role of two-pion\\ \>\> exchange with virtual delta-isobar excitation \\
12:45\>\>{\em End of Session}\\
13:00\>\>{\em Lunch}\\
{\em Early Afternoon Session}\\
        Chair: Johan Bijnens  \> \>  {\bf EFT in Atomic and Nuclear Physics}\\
14:20\>
        Hans-Werner Hammer  \>   Limit cycle physics \\
 \> (Seattle)\\
15:00\>
        Aurel Bulgac (Seattle)  \> What have we learned so far about dilute Fermi gases ? \\
15:40\>
        Lucas Platter (Bonn)  \>   An effective theory for the four-body system \\
16:05\>\>        
{\em Coffee}\\[2cm]
{\em Late Afternoon Session}\\
        Chair: Johan Bijnens    \> \>{\bf EFT in Nuclear Physics}\\
16:40\>
        Christoph Hanhart (J\"ulich)  \>     Subtleties in pion production reactions on \\\>\>few nucleon systems \\
17:05\>
        Hermann Krebs (Bonn)   \>  Neutral pion electroproduction off the deuteron \\
17:30\>
        Harald W. Grie{\ss}hammer    \>      Nucleon polarisabilities from Compton scattering\\ \> (Garching)\> off the proton and deuteron \\
17:55\> \>{\em End of Session}\\
18:30\> \>{\em Dinner}\\[3mm]
{\bf Thursday, December 16th, 2004}\\[3mm]
{\em Early Morning Session}\\
        Chair: Ulf-G. Mei\ss ner   \> \>{\bf  EFT in Particle Physics}\\
09:00\>
        Jan Stern (Orsay)    \>    Effective theory of Higgs-less electroweak
\\ \>\> symmetry breaking\\
09:40\>
        Johannes Hirn (Valencia)    \>     From mooses to 5D and back to large-Nc QCD ? \\
10:05\>
        Felix Sassen (J\"ulich)   \> Charm-strange mesons \\
10:30\>\>{\em Coffee}\\
{\em Late Morning Session}\\
        Chair: Ulf-G. Mei\ss ner\>  \>{\bf  Lattice QCD and ChPT}\\
11:00\>
        Martin Savage (Seattle)     \>     Lattice QCD and nuclear physics \\
11:40\>
        Silas Beane (Durham) \>    Nucleons and nuclei from lattice QCD \\
12:20\>
        Stephan D\"urr (Bern)    \>  Towards a lattice determinantion of NLO
\\ \>\> Gasser-Leutwyler coefficients \\
12:45\> \>{\em End of Session}\\
13:00\> \>{\em Lunch}\\
{\em Early Afternoon Session}\\
        Chair: Andreas Wirzba   \> \> {\bf Lattice QCD and ChPT}\\
14:20\>
        Karl Jansen (Zeuthen)  \>  Going chiral: overlap and twisted mass fermions \\
15:00\>
        Maarten Golterman     \>    Applications of ChPT to QCD with domain-wall \\ \>(San Francisco)\>fermions \\
15:40\> \>        {\em Coffee}
{\em Late Afternoon Session}\\
        Chair: Andreas Wirzba \> \> {\bf  Lattice QCD and ChPT}\\
16:30\>
        Christian B. Lang (Graz)   \>      Excited hadrons states from lattice calculations:\\ \>\> Approaching the chiral limit \\
17:10\>
        Hartmut Wittig (Hamburg)   \>      The epsilon-regime of QCD and its applications to\\ \>\> non-leptonic Kaon decays \\
17:45\>
        Thomas R. Hemmert  \>   Utilizing chiral effective field theory to understand\\ (Garching)\>\> lattice QCD simulations of baryon properties \\
18:20\> \>{\em End of Session}\\
18:30\> \>{\em Dinner}\\[20mm]
{\bf Friday, December 17th, 2004}\\[3mm]
Early Morning Session\\
        Chair: Ulf-G. Mei\ss ner \> \>   Lattice QCD and ChPT\\
09:00\>
        Elisabetta Pallante   \>       Progress, challenges and strategies in lattice QCD \\
 \> (Groningen)\\
09:40\>
        Jiunn-Wei Chen (Taipei)    \>      Lattice theory for low energy fermions at finite\\ \>\> chemical potential\\
10:20\> \>{\em Coffee}\\
{\em Late Morning Session}\\
        Chair: Ulf-G. Mei\ss ner \> \> {\bf  Lattice QCD and ChPT}\\
10:50\>
        Bugra Borasoy (Bonn)  \>   Finite volume effects using lattice
\\ \>\> chiral perturbation theory \\
11:15\>
        Christoph Haefeli (Bern)     \>    Finite volume effects for decay constants \\
11:35\>
        Helmut Neufeld (Wien)  \>  Isospin violation in semileptonic decays\\
12:00\>
        Johan Bijnens (Lund)  \> Farewell\\
12:10\> \>{\em Lunch and End of Workshop}\\
\end{tabbing}

\section{Participants and their email}

\begin{tabbing}
A very long namexxxxx\=a very long institutexxxxxx\=email\kill
S.\,R. Beane \> Univ.\ of New Hampshire \> silas@physics.unh.edu\\
V. Bernard \>LPT Strasbourg\> bernard@lpt6.u-strasbg.fr\\
J. Bijnens \> Lund Univ. \> bijnens@thep.lu.se \\
B. Borasoy \> Univ.\ Bonn\>borasoy@itkp.uni-bonn.de \\
P.\,C. Bruns \> Univ.\ Bonn\>bruns@itkp.uni-bonn.de\\
A. Bulgac \> Univ.\ of Washington\> bulgac@phys.washington.edu \\
P. B\"uttiker \> Univ.\ Bonn \>  buettike@itkp.uni-bonn.de \\
J.-W. Chen \> National Taiwan Univ. \> jwc@phys.ntu.edu.tw\\
G. Colangelo \> Univ.\ Bern \> gilberto@itp.unibe.ch\\
S. Descotes-Genon \>Univ.\ Paris-Sud \> sebastian.descotes@th.u-psud.fr\\
E. Dreisigacker \> WE-Heraeus-Stiftung \> dreisigacker@we-heraeus-stiftung.de\\
S. D\"urr \> Univ.\ Bern \> duerr@ifh.de\\
G. Ecker \> Univ.\ Wien\> gerhard.ecker@univie.ac.at \\
E. Epelbaum \> Jefferson Lab\> epelbaum@jlab.org\\
M. Frink \> Univ.\ Bonn \& FZ J\"ulich \> mfrink@itkp.uni-bonn.de \\
J. Gegelia \> Univ.\ Mainz \> gegelia@kph.uni-mainz.de\\
L. Girlanda \> ECT$^*$ Trento \> girlanda@ect.it\\
M. Golterman \> San Francisco State Univ. \>  maarten@stars.sfsu.edu\\
H.W. Grie{\ss}hammer \>  TU M\"unchen\> hgrie@ph.tum.de \\
C. Haefeli \> Univ.\ Bern \> haefeli@itp.unibe.ch\\
H.-W. Hammer \> Univ.\ of Washington\> hammer@phys.washington.edu\\
C. Hanhart \> FZ J\"ulich \> c.hanhart@fz-juelich.de \\
T.\,R. Hemmert \>TU M\"unchen\>  themmert@physik.tu-muenchen.de  \\
R.\,P. Hildebrandt \>TU M\"unchen\> rhildebr@ph.tum.de \\
J. Hirn \> Univ.\ de Valencia \> johannes.hirn@ific.uv.es  \\
B.\,R. Holstein \>Univ.\ of Massachusetts\> holstein@physics.umass.edu \\
K. Jansen \> DESY Zeuthen \> karl.jansen@desy.de \\
N. Kaiser \>TU M\"unchen\> nkaiser@ph.tum.de \\
M. Knecht \> CNRS-Luminy Marseille \> knecht@cpt.univ-mrs.fr\\
U. van Kolck \> Univ.\ of Arizona \> vankolck@physics.arizona.edu\\
H. Krebs \> Univ.\ Bonn \>  hkrebs@itkp.uni-bonn.de\\
S. Krewald \> FZ J\"ulich \> s.krewald@fz-juelich.de\\
B. Kubis \> Univ.\ Bonn \> kubis@itkp.uni-bonn.de\\
T. L\"ahde \> Lund Univ.\> talahde@thep.lu.se\\
C.\,B. Lang \> Univ.\ Graz \> christian.lang@uni-graz.at\\
V. Lensky \> FZ J\"ulich \> v.lensky@fz-juelich.de\\
H. Leutwyler  \> Univ.\ Bern \> leutwyler@itp.unibe.ch \\
A.\,V. Manohar \> UC San Diego\> manohar@ucsd.edu \\
U.-G. Mei{\ss}ner \> Univ.\ Bonn \& FZ J\"ulich \> meissner@itkp.uni-bonn.de\\
B. Moussallam \> Univ.\ Paris-Sud \> moussall@ipno.in2p3.fr\\
H. Neufeld \> Univ.\ Wien \> neufeld@ap.univie.ac.at\\
R. Ni{\ss}ler \> Univ.\ Bonn\> rnissler@itkp.uni-bonn.de \\
A. Nogga \> FZ J\"ulich \> a.nogga@fz-juelich.de\\
J.\,A. Oller \> Univ.\ de Murcia\> oller@um.es \\
E. Oset \> Univ.\ de Valencia \> oset@ific.uv.es\\
E. Pallante \> Univ.\ of Groningen \> pallante@phys.rug.nl\\
H.\,R. Petry \> Univ.\ Bonn \> petry@itkp.uni-bonn.de\\
A. Pich \> Univ.\ de Valencia \> antonio.pich@ific.uv.es\\
L. Platter \>Univ.\ Bonn \& FZ J\"ulich\> platter@itkp.uni-bonn.de\\
J. Prades \> Univ.\ de Granada \> prades@ugr.es\\
U. Raha \> Univ.\ Bonn \> udit@itkp.uni-bonn.de\\
E. Ruiz Arriola \> Univ.\ de Granada \> earriola@ugr.es\\
A. Rusetsky \>  Univ.\ Bonn \> rusetsky@itkp.uni-bonn.de\\
M.\,J. Savage \> Univ.\ of Washington \> savage@phys.washington.edu     \\
F. Sassen \> FZ J\"ulich \> f.sassen@fz-juelich.de\\
J. Stern \> Univ.\ Paris-Sud\> stern@ipno.in2p3.fr        \\
U.-J. Wiese \> Univ.\ Bern\> wiese@itp.unibe.ch \\
A. Wirzba \> FZ J\"ulich \> a.wirzba@fz-juelich.de\\
H. Wittig \> DESY Hamburg \> hartmut.wittig@desy.de \\
\end{tabbing}

\newabstract 
\begin{center}
{\large\bf $\mathbf{1/N}$ and Pentaquarks}\\[0.5cm]
Aneesh V. Manohar\\[0.3cm]
Department of Physics, University of California at San Diego,\\
La Jolla, CA 92093-0319, USA.\\[0.3cm]
\end{center}

The quantum numbers of exotic baryons states (those which are not $qqq$ states)
 can be obtained in the quark model, the Skyrme model, and in QCD using
 the $1/N_c$ expansion~\citetwo{JM1}{JM3}. The quantum numbers obtained using
 all three approaches agree. A quantum number, exoticness ($E$), is defined,
 and can be used to classify the states. The exotic baryons include the
 recently discovered $qqqq \bar q$ pentaquarks ($E=1$), as well as exotic
 baryons with additional $q \bar q$ pairs ($E >=1$). The mass formula for
 non-exotic and exotic baryons is given as an expansion in 1/N, and allows
 one to relate the moment of inertia of the Skyrme soliton to the mass of a 
constituent quark~\cite{JM1}.

Masses and widths of the flavor $\bf{27}$ and $\bf{35}$ pentaquark states in
 the same tower as the $\Theta^+$ are related by spin-flavor symmetry. The
 $\bf{27}$ and $\bf{35}$ states can decay within the pentaquark tower, as
 well as to normal baryons, and so have larger decay widths than the lightest
 pentaquark $\Theta^+$. The widths and branching ratios of the excited
 pentaquarks can be computed using the $1/N$ expansion~\cite{JM2}. 
An efficient operator method was developed, that greatly simplifies the 
computation of quark model matrix elements~\cite{AM1}, based on the coherent 
state picture for large $N$ baryons~\cite{AM2}.

The 1/N expansion also is applied to baryon exotics containing a single heavy
 antiquark. The decay widths of heavy pentaquarks via pion emission, and to
 normal baryons plus heavy $D^{(*)},B^{(*)}$ mesons are studied, and relations
 following from large-$N$ spin-flavor symmetry and from heavy quark symmetry
 are derived~\cite{JM2}.

\newabstract 
\begin{center}{\large \bf How well do we understand the interaction among the
    pions at low energies ?}   \\[0.5cm]
H. Leutwyler,\\ 
Institut f\"ur Theoretische Physik, Universit\"at Bern, CH-3012 Bern,
Switzerland\\[0.3cm]
\end{center}

S.~M.~Roy has shown long ago that, at low energies,
the $\pi\pi$ scattering amplitude can be expressed
in terms of twice subtracted dispersion integrals over physical region
imaginary parts. The crucial terms in his representation
are the two subtraction constants: at energies below $M_\rho$, the
poorly known contributions from the high energy part of the dispersion
integrals are very small. The fact that $\chi$PT accurately predicts the
values of the subtraction constants thus implies that, at low
energies, the entire scattering amplitude can now be calculated 
to a remarkable degree of precision: as compared to the information gathered
from phenomenology, the uncertainties of our results are typically smaller by
an order of magnitude. For a more detailed review and
references, see \citetwo{Mexico}{Colangelo Villasimius}. 

Then I discussed recent work \cite{ACCGL} on the pion matrix elements of the
scalar currents $\bar{u}u$, $\bar{d}d$, $\bar{s}s$ as well as the $K\pi$
transition matrix elements of $\bar{s}u$, $\bar{s}d$ and showed that the
singularities occurring near the $K\bar{K}$ threshold generate striking
features. Also, I reviewed the current state of knowledge of the scalar
radii and compared the low energy theorems for these with recent
experimental results. Lattice methods are now approaching the region of
light quark masses, where it becomes meaningful to analyze the data in
terms of the continuum effective theory. Although the available results
must yet be taken with a substantial grain of
salt, they do appear to confirm the simple picture that underlies our
crude theoretical estimates of 20 years ago. We can look forward to reliable
lattice measurements of several important effective couplings in the
foreseeable future. 

Yndur\'{a}in and  Pel\'{a}ez have criticized our work on
phenomenological grounds. I did not discuss
that, for lack of time, but refer the interested reader to
\citethree{Colangelo Villasimius}{ACCGL}{CCGL}.

\newabstract 
\begin{center}
{\large\bf Can one see the number of colors?} \\ [0.5cm]
U.-J.\ Wiese \\ [0.3cm]
Institute for Theoretical Physics, Bern University, CH-3012 Bern \\ [0.3cm]
\end{center}

It is well known that we live in a world with three quark colors. However, in 
contrast to textbook knowledge, some standard pieces of ``evidence'' for 
$N_c = 3$ do not at all imply three colors. The most prominent example is the
decay width $\Gamma(\pi^0 \rightarrow \gamma \gamma)$ which is proportional to
$[N_c (Q_u^2 - Q_d^2)]^2$. Standard textbooks assume the physical values of the
quark charges $Q_u = \frac{2}{3}$ and $Q_d = - \frac{1}{3}$ and conclude that 
only $N_c = 3$ is consistent with the experimentally observed width. However, 
in a world with $N_c$ colors, baryons consist of $N_c$ quarks. For general odd 
$N_c$ the baryons are fermions, and a proton consists of $\frac{1}{2}(N_c + 1)$
u-quarks and $\frac{1}{2}(N_c - 1)$ d-quarks, while a neutron contains 
$\frac{1}{2}(N_c - 1)$ u-quarks and $\frac{1}{2}(N_c + 1)$ d-quarks. In order 
to obtain the correct charges for proton and neutron, the quark charges must 
hence be adjusted to $Q_u = \frac{1}{2}(\frac{1}{N_c} + 1)$ and 
$Q_d = \frac{1}{2}(\frac{1}{N_c} - 1)$ which implies $N_c (Q_u^2 - Q_d^2) =1$ 
and thus eliminates the explicit $N_c$-dependence of 
$\Gamma(\pi^0 \rightarrow \gamma \gamma)$ [1]. Hence, the $\pi^0$ decay does
not at all imply $N_c = 3$. Remarkably, the above values of the quark charges 
also result from the anomaly cancellation conditions in a generalized standard 
model with arbitrary $N_c$ [1,2,3]. In particular, the cancellation of Witten's
global anomaly implies that $N_c$ must be odd.

At the level of chiral perturbation theory, for $N_f = 2$ flavors, the vertex 
for the $\pi^0 \rightarrow \gamma \gamma$ decay is contained in an 
$N_c$-independent Goldstone-Wilczek term. Indeed, for $N_f = 2$, it is 
impossible to infer the number of colors from low-energy experiments with pions
and photons [1]. For $N_f = 3$, the $\pi^0 \rightarrow \gamma \gamma$ decay 
gets a contribution from the Wess-Zumino-Witten term which is proportional to 
$N_c$. However, this contribution is completely cancelled by the 
$N_c$-dependent contribution from a Goldstone-Wilczek term which, in this case,
is proportional to $(1 - \frac{N_c}{3})$. Still, the $N_c$-dependence does not 
cancel in some other processes. For example, at tree level the width of the 
decay $\eta \rightarrow \pi^+ \pi^- \gamma$ is proportional to $N_c^2$ [1]. 
However, loop corrections turn out to be large, and this process seems not to 
be well suited for determining $N_c$ [4]. Due to mixing, the widths of the 
decays $\eta, \eta' \rightarrow \gamma \gamma$ have a complicated 
$N_c$-dependence [1]. Still, a loop calculation reveals that these processes 
indeed allow one to literally see the number of colors in low-energy 
experiments by detecting the photons emerging from the $\eta$ and $\eta'$ 
decays [5].

\ \newline
[1] O.\ B\"ar and U.-J.\ Wiese, Nucl.\ Phys.\ B609 (2001) 225. \newline
[2] S.\ Rudaz, Phys.\ Rev.\ D41 (1990) 2619. \newline
[3] A.\ Abbas, Phys.\ Lett.\ B238 (1900) 344; hep-ph/0009242. \newline 
[4] B.\ Borasoy and E.\ Lipartia, hep-ph/0410141. \newline
[5] B.\ Borasoy, Eur.\ Phys.\ J.\ C34 (2004) 317.

\newabstract 
\begin{center}
{\large\bf 
Chiral extrapolations on the lattice\\
with strange sea quarks
}\\[0.5cm]
S\'ebastien Descotes-Genon\\[0.3cm]
Laboratoire de Physique Th\'eorique,\\
B\^atiment 210, Universit\'e Paris-Sud/CNRS,\\
91405 Orsay, France\\[0.3cm]
\end{center}

The mass hierarchy among light quarks indicates that
the (light but not-so-light) strange quark may 
play a special role in the low-energy dynamics of QCD.
In particular, strange sea-quark pairs may induce significant
differences between the patterns of chiral symmetry breaking
in the chiral limits of $N_f=2$ massless flavours ($m_u=m_d=0$,
$m_s$ physical) and $N_f=3$ ($m_u=m_d=m_s=0$)~\cite{param}.
Such a difference, related to a significant violation 
of the Zweig rule in the scalar sector, would lead to a value of
the $N_f=3$ quark condensate much smaller than its $N_f=2$ counterpart, and 
thus to instabilities in three-flavour chiral series by suppressing
leading-order terms and enhancing numerically next-to-leading-order 
terms~\cite{resum}. Indirect dispersive estimates suggest that
this damping effect could be significant~\cite{uuss}.

I focus on how unquenched lattice
simulations with three dynamical flavours
could be sensitive to this suppression~\cite{extrapol}. I explain how chiral 
extrapolations of masses and decay constants of the pion and kaon could 
be affected by the presence of massive $s\bar{s}$-pairs in the sea, and I 
estimate finite-size effects related to the periodic boundary conditions 
imposed in lattice simulations~\cite{finvol}. The impact of strange 
sea-quark pairs on three-flavour chiral symmetry breaking could be assessed
through the quark-mass dependence of two ratios based on 
$F_\pi^2M_\pi^2$ and $F_K^2M_K^2$ and affected by only small
finite-volume corrections for $L \sim$ 2.5 fm.

\newabstract 
\begin{center}

{\large\bf A Linearized GDH Sum Rule}\\[0.5cm]

Barry R. Holstein\\[0.3cm]
Department of Physics-LGRT\\
University of Massachusetts\\
Amherst, MA  01003\\[0.3cm]

\end{center}

The GDH sum rule relates a particle's anomalous magnetic moment
$\kappa$ to a weighted integral over polarized Compton scattering
cross sections\cite{gdh}---
\begin{equation}
{2\pi\alpha\over M^2}\kappa^2={1\over \pi}\int_0^\infty
{d\omega\over \omega}\Delta\sigma_s(\omega)
\end{equation}
where
$$\Delta\sigma_s(\omega)=\sigma_{1+s}(\omega)-\sigma_{1-s}(\omega)$$
This has been well tested and seems to work at least in the case
of the proton.  In the case of QED, however, since $\kappa$ itself
is ${\cal O}(\alpha)$ one requires the cross sections at one loop
level to get the well-known Schwinger result---$\kappa=\alpha/
2\pi$.  In addition, at ${\cal O}(\alpha)$ this leads to a
consistency condition, as pointed out by Altarelli, Cabibbo, and
Maiani\cite{acm}.

Together with Vladimir Pascalutsa and Marc Vanderhaeghen we have
developed a modified GDH sum rule that obviates this
situation\cite{gdh1}. The idea is to add a Pauli moment $\kappa_0$
to the calculation and to calculate a derivative of the relevant
cross sections with respect to this Pauli moment, which is then
set to zero.  The resulting sum rule reads (at lowest order)
\begin{equation}
{4\pi\alpha\over M^2}\kappa={1\over \pi}\int_0^\infty
{d\omega\over \omega}\Delta\sigma_1(\omega)
\end{equation}
and is now linear in $\kappa$, where $\Delta\sigma_1$ is the
derivative of the cross section at zero Pauli moment.  This
allows, for example, the Schwinger moment to be found from tree
level input. In the case of pion-nucleon scattering this allows a
simple calculation of the one loop magnetic moment of the proton
or neutron to all orders in $m_\pi$. Possible future applications
could include a two loop calculation of the magnetic moment using
one loop input.

\newabstract 
\begin{center}
{\large\bf Determination of the $\pi N$ scattering lengths 
from the experiments on pionic deuterium}\\[0.5cm]
Ulf-G. Mei{\ss}ner$^{1,2}$, Udit Raha$^1$ and {\bf Akaki Rusetsky}$^{1,3}$\\[0.3cm]
$^1$Universit\"{a}t Bonn, Helmholtz-Institut f\"{u}r
Strahlen- und Kernphysik (Theorie),\\
Nu\ss allee 14-16, D-53115 Bonn, Germany
\\[0.3cm]
$^2$Forschungszentrum J\"{u}lich, Institut f\" {u}r Kenphysik (Theorie),\\
D-52425 J\"{u}lich, Germany
\\[0.3cm]
$^3$On leave of absence from: High Energy Physics Institute,\\
Tbilisi State University, University St.~9, 380086
Tbilisi, Georgia
\\[0.3cm]
\end{center}

We use the framework of effective field theories  to discuss the determination
of the $S$-wave $\pi N$ scattering lengths $a_+$ and $a_-$ from the recent 
high-precision measurements of pionic deuterium observables~\cite{PSI}. 
Initially, the precise value of the pion-deuteron scattering length 
$a_{\pi d}$ is extracted from the experimental data. Next, $a_{\pi d}$ is 
related to the $S$-wave $\pi N$ scattering lengths in the multiple-scattering 
series, which are derived in the so-called Heavy Pion effective field 
theory. We discuss the use of the information, coming from pionic deuterium,
for constraining the values of the $\pi N$ scattering lengths in the full 
analysis, which also includes the input from the pionic hydrogen energy shift 
and width measurements. In particular, we throughly investigate the accuracy limits for 
this procedure and give a detailed comparison to other effective field 
theory approaches~\citethree{Meissner}{Borasoy}{Beane},
as well as with the earlier 
work on the subject, carried out within the potential framework. The present
results will be published elsewhere~\cite{deuteron}.

\newabstract 
\begin{center}{\Large{\bf Spectrum and decays of the Kaonic Hydrogen
}}\\
\bigskip
{U.-G. Mei\ss ner, {\bf U. Raha}, and A. Rusetsky}\\
\bigskip
{\em Helmholtz-Institut f\"ur Strahlen- und Kernphysik, Universit\"at Bonn}\\
{\em Nu\ss allee 14-16, 53115 Bonn, Germany}
\end{center}

Recent accurate measurements \citetwo{Baldini}{Cargnelli} of the strong energy shift and the lifetime
of the ground state of kaonic hydrogen by DEAR collaboration at
LNF-INFN
allow one to extract the precise values of the $KN$ scattering lengths
from
the data. To this end, one needs to relate the latter quantities to
the observables of the kaonic hydrogen at the accuracy that matches the
experimental precision.
In our recent investigations, the problem is considered within the
non-relativistic effective Lagrangian approach, which has been previously
used
to describe the bound $\pi^+\pi^-$ , $\pi^-p$ and $\pi K$  systems (see, e.g. 
\citefour{Gall}{Lyubovitskij}{Zemp}{Schweizer}). We obtain \cite{Meissner} a
general
expression of the strong shift of the level energy and the decay width in
terms
of the $KN$ scattering lengths, at $\mathcal{O}(\alpha,m_d-m_u)$ as compared to the
leading-order result. It is shown that,
due to the presence of the unitarity cusp in the $K^-p$
elastic scattering amplitude above threshold, the isospin-breaking
corrections turn out to be very large. This, however, does not affect
the accuracy of the extraction of the scattering lengths from the
experiment.

\newabstract 
\begin{center}
{\large\bf Anomalous decays of $\mbox{\boldmath$\eta$}$ and 
$\mbox{\boldmath$\eta'$}$ with coupled channels}\\[0.5cm]
B.\ Borasoy and {\bf R.\ Ni{\ss}ler}\\[0.3cm]
Helmholtz-Institut f\"ur Strahlen- und Kernphysik,  \\
Universit{\"a}t Bonn, 53115 Bonn, Germany.\\[0.3cm]
\end{center}

The axial anomaly of QCD dominates various low-energy processes involving
light mesons. Interesting examples for such anomalous processes are the
decays of $\pi^0$, $\eta$ and $\eta'$ into two photons and the decays
$\eta, \eta' \to \pi^+ \pi^- \gamma$, since they reveal information on
both the chiral and---involving the $\eta'$---the axial $U(1)$ anomaly.

An analysis of the experimental data of the anomalous $\eta$ and $\eta'$ decays
demands for the inclusion of resonances in the theoretical framework.
To this end we combine the effective $U(3)$ chiral Lagrangian with a coupled channel
Bethe-Salpeter equation which satisfies unitarity constraints
and generates vector-mesons from composite states of two pseudoscalar mesons.
The phase shifts of meson-meson scattering, particularly
in the important $\rho$-resonance channel, are precisely reproduced.

The resulting $T$ matrix for meson-meson rescattering is then used as an 
effective vertex which is included in the loop calculation of the decay 
amplitudes.  This procedure manifestly preserves electromagnetic gauge 
invariance and exactly matches to the amplitudes obtained in one-loop 
Chiral Perturbation Theory.
Furthermore, in the chiral limit our approach naturally reproduces the results
dictated by the chiral anomaly of QCD. Overall good agreement with the 
available experimental spectra and decay parameters is obtained
 \citetwo{BN1}{BN2}.

Contrary to the common picture of Vector Meson Dominance, where the 
simultaneous exchange of two vector mesons is the only contribution to the 
two-photon decays with both photons being off-shell, the corresponding 
diagram in our approach is highly suppressed with respect to the excitation 
of only one $p$-wave resonance. Therefore, the resulting transition form 
factor in the process $\eta' \to \gamma^* \gamma^*$ is significantly smaller. 
This prediction may be verified at the planned WASA@COSY facility and is also 
of interest for the determination of the hadronic contribution to the 
anomalous magnetic moment of the muon.

\newabstract 
\begin{center}
{\large\bf Radiative $K_{e3}$ decays revisited}\\[0.5cm]
J\"urg Gasser$^1$, {\bf Bastian Kubis}$^{1,2}$,
  Nello Paver$^3$ and Michela Verbeni$^4$\\[0.3cm]
{\footnotesize
$^1$Institut f\"ur theoret.\ Physik, Universit\"at Bern, 
  Sidlerstrasse 5, CH-3012 Bern, Switzerland \\[0.2cm]
$^2$HISKP (Abt.\ Theorie), Universit\"at Bonn, 
  Nussallee 14-16, D-53115 Bonn, Germany \\[0.2cm]
$^3$Dip. di Fis.\ Teor., Univ.\ di Trieste, and INFN-Trieste, 
  Str.\ Costiera 11, I-34100 Trieste, Italy \\[0.1cm]
$^4$Depto.\ F\'isica Te\'orica y del Cosmos,
  Univ.\ de Granada, E-18002 Granada, Spain
}
\end{center}

We have studied~\cite{GKPV} the process
$K_L \to \pi^\mp e^\pm \nu_e \gamma ~ [ K_{e3\gamma} ]$
by combining two  theoretical concepts: 
Low's theorem yields the bremsstrahlung amplitude in terms of the
$K_{e3}$ form factors $f_+$ and $f_2$ ~\cite{FFS}, 
while Chiral Perturbation Theory (ChPT) at one
loop~\cite{BEG} and beyond~\cite{GKPV} 
provides an assessment of the structure dependent 
contributions. 

For the ratio of radiative decay rate (with cuts on photon energy,
$E_\gamma^*>30\,$MeV, and photon-electron angle, $\theta_{e\gamma}^*>20^\circ$) 
to the non-radiative rate, we find 
\[
R = \Gamma(K_{e3\gamma})/\Gamma(K_{e3})
  = \left(0.96 \pm 0.01\right) \times 10^{-2} ~~. 
\]
This value is very insensitive to the precise shape of $f_+$, while the effect
of $f_2$ is suppressed by powers of $m_e$ and can be neglected. 
Structure dependent terms contribute less than 1\% to $R$.  Logarithmically
enhanced effects from radiative corrections have been included, the 
remaining electromagnetic corrections dominate the 
uncertainty quoted.  This prediction agrees very well with a recent
measurement by NA48~\cite{na48}, but is at variance
with results from KTeV~\citetwo{ktev}{ktev_04}.

Structure dependent terms can be accessed experimentally via differential
distributions.  
We have demonstrated that $d\Gamma/dE_\gamma^*$ is sensitive to
essentially only one linear combination of structure functions. 
In ChPT, these  are real and constant to
a reasonable approximation, and we find, for the relevant combination,
\[
C' = -1.6 \pm 0.4 ~~,
\]
to be compared with the experimental result $C'=-2.5^{+1.5}_{-1.0}\pm
1.5$ \cite{ktev}.

\newabstract 
\begin{center}
{\large\bf Electromagnetic LEC's and QCD n-point 
functions resonance models}\\[0.5cm]
B. Ananthanarayan$^1$, {\bf B. Moussallam}$^2$\\[0.3cm]
$^1$ Centre for High Energy Physics,
IISc, Bangalore, 560012,India\\[0.3cm]
$^2$IPN, Universit\'e Paris-Sud, 91406 Orsay C\'edex, France\\
[0.3cm]
\end{center}
The search for approximate estimates of the chiral couplings (LEC's) 
is important for improving the predictivity and for a better 
understanding of the workings of the chiral expansion. The strategy is to  
establish relations with QCD Green's functions for which approximations
may be built based on the large $N_c$ expansion  and constraints from 
low as well as large energies\cite{eglpr}. 
The chiral-electromagnetic LEC's are linked to QCD correlators via convolution 
integrals\cite{mouss97}. Convergence is controlled by the QED 
counterterms, implying that some 
electromagnetic LEC's depend on the short distance 
renormalization scale $\mu_0$. As an application, an answer to the question 
raised in ref.\cite{grs} of defining
a ``pure QCD'' quark mass $\overline{m}_f(\mu,\mu_0)$ in terms of the
physical quark mass $m_f(\mu_0)$ is provided by
\begin{equation}
\overline{m}_f(\mu,\mu_0)= m_f(\mu_0) \big(
1+ 4e^2 {Q_f^2} ({K^r_9(\mu,\mu_0) + K^r_{10}(\mu,\mu_0)})  \big)\ ,
\end{equation}
and $K_9+K_{10}$ could be determined model independently using input
from $\tau$ decays.
A number of LEC's are related
to QCD 4-point functions which we have recently investigated\cite{am04}. 
We consider a large $N_c$ modeling starting from the resonance Lagrangian 
of ref.\cite{eglpr} in which resonances maybe ascribed a chiral order and
which contains all terms of order $p^4$. At this level a number of asymptotic
constraints fail to get fulfilled, the resonance Lagrangian is then extended
to include a set of terms of order $p^6$, restricting ourselves to terms
of the form $RR'\pi$. The asymptotic constraints can then be obeyed and this
gives rise to a set of Weinberg-type equations. It turns out that
all the needed resonance coupling constants get determined in this way.

\newabstract 
\begin{center}
{\large\bf Final State Interactions in Hadronic $D$ Decays}\\[0.5cm]
Jos\'e A. Oller\\[0.3cm]
Departamento de Física, Universidad de Murcia,\\
E-30071 Murcia, Spain\\[0.3cm]
\end{center}

We  study the final state interactions of the three light pseudoscalars 
produced in the D decays, $D_s^+\rightarrow \pi^- \pi^+ \pi^+$ \cite{prl1}, 
$D^+\rightarrow  \pi^- \pi^+ \pi^+$ \cite{prl2} and 
$D^+\rightarrow K^- \pi^+ \pi^+$ \cite{prl3}, where 
high statistically significant evidences for the light scalar resonances 
$\kappa$ or $K^*0(800)$ and $\sigma$ or $f_0(600)$ are
obtained from high statistics data \citethree{prl1}{prl2}{prl3} with dominating S-wave 
dynamics. We discuss why in 
these reactions the 
$\kappa$ and $\sigma$ poles are clearly visible as broad bumps in contrast 
with the 
scattering amplitudes, where one can only see slowly increasing amplitudes 
as shoulders. Regarding 
the final state interactions of the producing pseudoscalars, we  
assume as in the experimental analyses refs.\citethree{prl1}{prl2}{prl3} the
 spectator approach in terms of Bose-symmetrized amplitudes. However,
  we are able to show that the large 
final state interactions corrections present in the data are compatible 
with the pseudoscalar-pseudoscalar two body scattering amplitudes. 
This aspect was an unsolved 
 issue by the experimental analyses that used a coherent sum of 
 Breit-Wigner's, following the isobar model, which did not give rise to
 the same phases as those measured in scattering experiments. Similar 
problems also appear in B decays, see  refs.\cite{meiss} for earlier and 
related partial discussions.  Our agreement with the rather precise data 
on D to $3\pi$ and $K 2\pi$, where S-wave dynamics dominate, 
also implies an interesting 
experimental confirmation of the S-wave T-matrices obtained from unitary 
CHPT \citetwo{npa}{jamin}.

\newabstract 
\begin{center}
{\large\bf Chiral dynamics of baryon resonances}\\[0.5cm]
E. Oset\\
{\small Departamento de F\'{\i}sica Te\'orica and IFIC,
Centro Mixto Universidad de Valencia-CSIC,} \\
{\small Institutos de
Investigaci\'on de Paterna, Aptd. 22085, 46071 Valencia, Spain}\\[0.3cm]
\end{center}
One of the interesting topics in Hadron Physics is the 
dynamical generation of baryon resonances using tools of unitary chiral
perturbation theory. Admitting that most of the resonances can approximately
qualify as made of three constituent quark states, some of them are better
represented as kind of molecular states of a meson and a baryon, and using
 unitarized versions of chiral perturbation theory one can make a systematic 
 study of these states.A study of poles of the scattering matrix of
coupled channels in the interaction of the octet of baryons of $1/2^+$  with
the octet of mesons of  $0^-$ was shown in \cite{cola} to lead to two octets 
and one singlet of
dynamically generated resonances, most of which correspond to well known
resonances of $1/2^-$, while there is a prediction for new resonances.
 Particularly it was found that there are two states close to the
nominal $\Lambda(1405)$. Some reactions where these two states could be
disentangled are the $K^-p \to \gamma \Lambda(1405)$ \cite{nacher} and $\gamma p
\to K^* \Lambda(1405)$ \cite{hyodo}.

  On the other hand the interaction of the decuplet of baryons of the $\Delta$
with the octet of pseudoscalar mesons also leads to  a group of well known
resonances, while there are predictions for new ones \cite{sarkar}.
 One of the striking examples of this latter case is a resonance formed with
  the interaction of $\Delta  K$ in isospin I=1 \cite{delka}.  This resonance
  leads to large cross sections for $\Delta  K$ in I=1, compared to the case with
  I=2, which are amenable of experimental search in some pp collisions leading
  to $\Delta  K$ in the final state.

\newabstract 
\begin{center}
{\large\bf The Dalitz Decay $\pi^0\to e^+e^-\gamma$}\\[0.5cm]
Karol Kampf$^1$, {\bf Marc Knecht}$^2$ and Ji\v{r}\'{\i} Novotn\'y$^1$\\[0.3cm]
$^1$Institute of Particle and Nuclear Physics, Charles
University,\\
 V Holesovickach 2, 180 00 Prague 8, Czech Republic\\[0.3cm]
$^2$Centre de Physique Th\'eorique\footnote{Unit\'e mixte de
recherhce (UMR 6207) du CNRS et des universit\'es Aix-Marseille I,
Aix-Marseille II, et du Sud Toulon Var; laboratoire affili\'e \`a
la FRUMAM (FR 2291).}, CNRS-Luminy, Case 907,\\
 F-13288
Marseille Cedex 9, France\\[0.3cm]
\end{center}

The ${\cal O}(\alpha)$ corrections to the Dalitz decay 
$\pi^0\to e^+e^-\gamma$ have been studied \cite{dalitz}
in the framework of two-flavour chiral perturbation theory 
with virtual photons and leptons \cite{Knecht:1999ag}.
Besides the fact that it constitutes the second
most important decay channel of the pion, the Dalitz decay gives
access to the off-shell $\pi^0 - \gamma^* - \gamma^*$
transition form factor 
${\cal A}_{\pi^0  \gamma^*  \gamma^*}(q_1^2,q_2^2)$ at low
energies, and in particular to its slope parameter $a_\pi$, 
defined as 
${\cal A}_{\pi^0  \gamma^*  \gamma^*}(q^2,0)~=~1~+~a_\pi(q^2/M_{\pi^0}^2~+~\cdots$. 
We have not only considered the ${\cal O}(e^5)$ one photon reducible 
corrections to the leading, ${\cal O}(e^3)$, amplitude, but also the
NLO corrections generated by the one fermion reducible and the 
one particle irreducible contributions to the Dalitz plot distribution,
$
\frac{d\Gamma}{dx dy}\,=\, 
[1 + \delta(x,y)]\,\frac{d\Gamma^{LO}}{dx dy}.
$
We have used a large-$N_C$ inspired representation \cite{Knecht:2001xc}
of the form factor ${\cal A}_{\pi^0  \gamma^*  \gamma^*}(q_1^2,q_2^2)$, 
based on its known chiral and QCD short distances properties.

Although their contribution to the total decay rate is very mall,
we have found that the one particle irreducible ${\cal O}(e^5)$ corrections
are sizeable in some region of the Dalitz plot. Within this framework,
we obtain the following prediction for the slope parameter:
$
a_\pi = 0.034 \pm 0.005,
$
which agrees well with the value obtained from (model dependent) 
extrapolations of the form factor ${\cal A}_{\pi^0  \gamma^*  \gamma^*}(q^2,0)$
measured at higher energies. Omission of the ${\cal O}(e^5)$ one photon
irreducible radiative corrections in the extraction of $a_\pi$ from
the experimental Dalitz plot distribution, 
$d\Gamma^{exp}/dx - \delta_{QED}(x)\,(d\Gamma^{LO}/dx)
= (d\Gamma^{LO}/dx)[1 + a_\pi x]$, would decrease this experimental
determination of $a_\pi$ by $0.005$.

\newabstract 
\begin{center}
{\large\bf The pion vector form factor and the muon
    anomalous magnetic moment}\\[0.5cm]
Gilberto Colangelo \\[0.3cm]
Institut f\"ur Theoretische Physik der Universit\"at Bern \\
Sidlerstr. 5, 3012 Bern, Switzerland
\end{center}
The hadronic vacuum polarization contribution to $a_\mu$ is dominated by
the $\pi \pi$ contribution (see, e.g. \cite{DEHZ}), which is given by the
pion vector form factor. The Omn\'es representation expresses the latter in
terms of its phase which, below 16 $M_\pi^2$ exactly, and up to 4 $M_K^2$
to a good approximation, is equal to the $\pi \pi$ P-wave phase shift.
The latter is strongly constrained by analyticity, unitarity and chiral
symmetry \cite{CGL}. We call the Omn\'es function constructed with the $\pi
\pi$ P-wave phase shift $G_1(s)$,
and represent the physical vector pion
form factor as a product of three terms:
\begin{equation}
F_\pi(s)= G_1(s) G_2(s) G_\omega(s)
\label{eq1}
\end{equation}
where $G_2(s)$ accounts for inelastic effects, and $G_\omega(s)$ takes into
account the contribution of the $\omega$ through its interference with the
$\rho$. Both functions $G_2$ and $G_\omega$ can be described in terms of a
small number of parameters which are fixed by fitting the data. Moreover, 
the phase of the inelastic contribution is constrained by an inequality due
to Lukaszuk \cite{Eidelman:2003uh}. 

I have presented preliminary results (see also \cite{Colangelo:2003yw})
for a calculation of the hadronic vacuum polarization contribution to
$a_\mu$ below 1 GeV obtained on the basis of Eq.~(\ref{eq1}) and on the
analysis of $e^+e^-$ data \citetwo{Akhmetshin:2003zn}{Aloisio:2004bu}.

\newabstract 
\begin{center}
{\large\bf Partially Quenched Chiral Perturbation Theory at NNLO}\\[0.5cm]
Johan Bijnens$^1$ and {\bf Timo L\"ahde}$^1$\\[0.3cm]
$^1$Department of Theoretical Physics, Lund University,\\
S\"olvegatan 14A, SE - 22362 Lund, Sweden\\[0.3cm]
\end{center}

For computational reasons, the inclusion of the fermionic determinant, 
or the sea quark effects, into Lattice QCD simulations is still impractical
for sea quark masses that are close to the physical $u,d$ quark masses.
However, recent progress has made Lattice QCD simulations with sea quark 
masses of a few tens of MeV available, a situation which is referred to as
partially quenched~(PQ) QCD. The simulation results then have to be
extrapolated to the physical quark masses using Chiral Perturbation 
Theory~($\chi$PT). The generalization of $\chi$PT to the quenched case 
(without sea quarks) or to the partially quenched case (sea quark masses 
different from the valence ones) has been carried out by Bernard and 
Golterman in Refs.~\cite{BG}. The quark mass dependence of partially quenched
chiral perturbation theory (PQ$\chi$PT) is explicit, and thus the
limit where the sea quark masses become equal to the valence quark
masses can be taken. As a consequence, $\chi$PT is included in
PQ$\chi$PT and the free parameters, or low-energy constants (LEC:s),
of $\chi$PT can be directly obtained from those
of PQ$\chi$PT~\citetwo{BG}{Sharpe}.

The calculation of charged pseudoscalar meson masses and decay
constants to one loop (NLO) in PQ$\chi$PT has been carried out
in Refs.~\citetwo{BG}{Sharpe}, and first results for the mass of a 
charged pseudoscalar meson at two loops, or next-to-next-to-leading order 
(NNLO) in PQ$\chi$PT, may be found for degenerate sea quark masses
 in Ref.~\cite{BDL}. The NNLO result for the decay constants of the charged
pseudoscalar mesons in three-flavor PQ$\chi$PT has recently appeared in
Ref.~\cite{BL1}. The need for such calculations is clear as NNLO
effects have already been detected in Lattice QCD 
simulations~\cite{Latt}. A calculation of the pseudoscalar
meson masses for nondegenerate sea quarks is in progress.

\newabstract 
\begin{center}
{\large\bf Effective Lagrangians in the Resonance Region}\\[0.5cm]
Antonio Pich\\[0.3cm]
Departament de F\'{\i}sica Te\`orica, IFIC, Universitat de Val\`encia -- CSIC,\\
Apt. Correus 22085, E-46071 Val\`encia, Spain\\[0.3cm]
\end{center}

In the resonance region chiral perturbation theory is no longer valid and one
must introduce a different effective field theory with explicit massive fields.
Chiral symmetry still provides stringent dynamical constraints, but the
usual chiral power counting breaks down in the presence of higher energy scales.
The limit of an infinite number of quark colours 
provides an alternative power
counting to describe the meson interactions \cite{PI:02}.
Assuming confinement, the strong dynamics at large $N_C$ is given by tree
diagrams with infinite sums of hadron exchanges, which correspond
to the tree approximation to some local effective lagrangian
\citetwo{HO:74}{WI:79}.
Hadron loops generate corrections suppressed by factors of $1/N_C$.

The large $N_C$ limit of the ``Resonance Chiral Theory''  \cite{EGPR:89}
has been investigated in many works and a very successful leading
order phenomenology already exists 
\citesix{MO:95}{KPdR}{KN:01}{RPP:03}{CEEPP:04}{BGLP:03}.
More recently, the problems associated with quantum corrections involving 
heavy resonance propagators have been investigated  \cite{RSP:04}. 
This constitutes
a first step towards a systematic procedure to evaluate next-to-leading order
contributions in the $1/N_C$ counting.

\newabstract 
\begin{center}
{\large\bf Infrared regularization for spin-1 fields}\\[0.5cm]
{\bf Peter C. Bruns}$^{1}$ and {Ulf-G. Mei\ss ner}$^{1,2}$\\[0.3cm]
$^1$Univ. Bonn, HISKP (Th), D-53115 Bonn, Germany\\[0.3cm]
$^2$FZ J\"ulich, IKP (Th),  D-52425 J\"ulich, Germany\\[0.3cm]
\end{center}

\noindent
We consider chiral perturbation theory with explicit spin-1
degrees of freedom (vector and axial-vector mesons), utilizing the
antisymmetric tensor field formulation.
When vector mesons appear in loops, the appearance of the large mass
scale (the vector/axial-vector meson mass in the chiral limit)
complicates the power counting.
In essence, loop diagrams pick up large contributions when the 
loop momentum is close to the vector meson mass. To the
contrary, the contribution from the soft poles (momenta of the order of the
pion mass) that leads to the interesting chiral terms of the low-energy
 EFT (chiral logs and alike) obeys power counting. The standard case 
of infrared regularization \cite{BL}, 
where the heavy particle line is conserved in the (one-loop) graphs,
can only be applied to a subset of interesting loop graphs with vector mesons.
In the case of spin-1 fields, new classes of self-energy graphs
appear. The case for lines with small external momenta but a vector meson
line appearing inside the diagram  was analyzed in \cite{BM} and  the 
infrared singular part for such types of integrals was explicitly constructed.
As explicit examples, the Goldstone boson self-energy and the triangle 
diagram are worked out.
As an application, we consider the pion mass dependence of the $\rho$-meson
mass $M_\rho$. We show that although there are many contributions with unknown
low-energy constants, still one is able to derive a compact formula 
for the pion (quark) mass dependence of $M_\rho$. 
We analyze existing lattice data \cite{CPPACS} and conclude that 
the  $\rho$-meson mass in the
chiral limit is bounded between 650 and 800~MeV. We have also discussed the
$\pi\rho$ sigma term.

\medskip\noindent
Work supported in part by DFG, SFB/TR-16 ``Subnuclear Structure of Matter''.

\newabstract 

\begin{center}
{\large\bf A Large N$_c$ Hadronic Model}\\[0.5cm]
{Johan Bijnens$^{1}$, Elvira G\'amiz$^{2}$,
and {\bf Joaquim Prades}$^3$}\\[0.5cm]
$^{1}$ Department of Theoretical Physics 2, Lund University\\
S\"olvegatan 14A, S 22362 Lund, Sweden.\\[0.5cm]

$^{2}$ Department of Physics \& Astronomy, University of Glasgow\\
Glasgow G12 8QQ, United Kingdom.\\[0.5cm]

$^{3}$ Departamento de
 F\'{\i}sica Te\'orica y del Cosmos, Universidad de Granada\\
Campus de Fuente Nueva, E-18002 Granada, Spain.\\[0.5cm]
\end{center}

We present a large $N_c$ hadronic model with a single hadronic
state per channel and construct
all two, three-point  and some four-point
Green's functions in the chiral limit.
These Green's functions contain all the constraints from CHPT
at NLO, the large $N_c$ structure, and the maximum of the short-distance
constraints, namely, OPE and Brodsky-Lepage-like (or quark-counting-rule) ones.
We point out a general problem for large $N_c$ hadronic 
models that prevents
from imposing all those short-distance constraints on three-point
and higher Green's functions with a finite number of resonances.
This problem also exists for Green's functions that vanish to all orders
in massless perturbative QCD. 
We also give the complete result
outside the chiral limit for two-point functions, and some
three-point ones in the chiral limit.  Three-point functions
outside the chiral limit are under way \cite{BGP1}. We also comment on the
 application  to calculate the $\hat B_K$ parameter in the chiral limit
for which we present preliminary results  and outside 
the chiral limit \cite{BGP2}

\newabstract 
\begin{center}
{\large\bf Baryon masses in  CHPT and from the lattice}\\[0.5cm]
{\bf Matthias Frink}$^{1,2}$, {Ulf-G. Mei\ss ner}$^{1,2}$ and Ilka Scheller$^1$\\[0.3cm]
$^1$Univ. Bonn, HISKP (Th), D-53115 Bonn, Germany\\[0.1cm]
$^2$FZ J\"ulich, IKP (Th),  D-52425 J\"ulich, Germany\\[0.1cm]
\end{center}
We analyze the masses of the ground state octet baryons in a version
of cut-off regularized and in dimensionally regularized  baryon CHPT 
at third, improved third and fourth order 
in the chiral expansion \cite{SFM}, 
based on  Ref.\cite{BHM} (and using earlier calculations of baryon masses
\citetwo{BM}{FM}).  The corresponding LECs are
determined by the conditions that the nucleon mass $m_N$ takes its physical 
value for the physical values of $M_\pi$ and $M_K$ and that one 
obtains a fair description
of the MILC data \cite{MILC} already at third order including the improvement term.

\vspace{-0.38cm}

\begin{figure} [h]
 \centering
 \begin{minipage}[c]{.45\textwidth}
{
Shown is the pion mass dependence of the nucleon mass 
in dimensional regularization at third (dashed), improved third
  (solid) and fourth (dot-dashed) order, respectively. The dotted line
  represents the fourth order calculation from \protect\cite{BM}.
The 
three flavor data (staggered fermions) are from 
the MILC collaboration \protect\cite{MILC}.
The filled circle gives $m_N = 940\,$MeV at $M_\pi = 140\,$MeV.
}
 \end{minipage}
 \hfill
 \begin{minipage}[c]{.50\textwidth}
   \centering
   \includegraphics [scale=.28,angle=270]{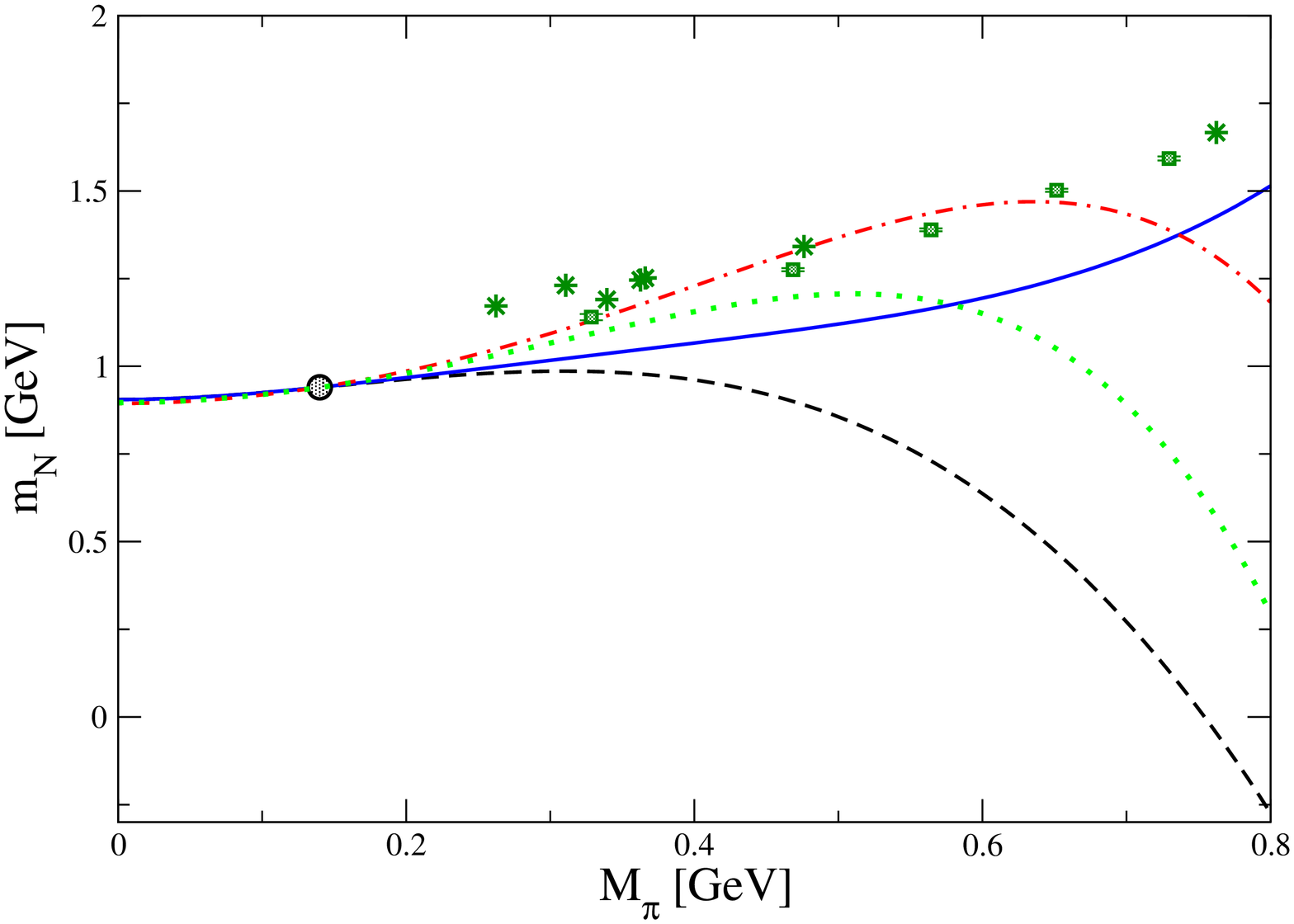}
 \end{minipage}
\end{figure}

\vspace{-0.5cm}

\noindent
From the kaon mass dependence of $m_N$, we deduce that the octet mass in the
chiral limit lies in the range from 770~MeV to 1070~MeV. We have also given
extrapolation functions for the $\Lambda$, the $\Sigma$ and the $\Xi$ and
compared to the MILC data. We find that the chiral extrapolation functions
(with all parameters fixed from the nucleon mass) 
are flatter than what is indicated by these data.

\newabstract 
\begin{center}
{\large\bf Chiral dynamics in few--nucleon systems}\\[0.5cm]
Evgeny Epelbaum$^1$\\[0.3cm]
$^1$Jefferson Laboratory, Theory Division, Newport News, VA 23606, USA.\\[0.3cm]
\end{center}

Chiral Effective Field Theory (EFT) has become a standard tool for analyzing the properties 
of hadronic systems at low energy in a systematic and controlled way 
based upon the approximate and spontaneously broken chiral symmetry of Quantum Chromodynamics 
(QCD). Based on Weinberg's idea to use chiral EFT to derive nuclear forces, 
this method has also been successfully applied to few--nucleon problems 
\citetwo{weinb}{ord}. 
Energy--independent and hermitian nuclear potentials can be derived from the chiral Lagrangian
e.g.~using the method of unitary transformation \cite{EGM1}. This scheme can also be applied to 
derive the corresponding nuclear current operator.  We have used this framework to study 
the 2N system at next--to--next--to--next--to--leading order (N$^3$LO) in the chiral expansion \cite{EGMn3},
see also \cite{EM}.
The theoretical uncertainty for scattering observables at N$^3$LO 
is expected to be of the order $\sim$ 0.5\%, 7\% and 25\% at laboratory energy $\sim$ 50, 150 and 250 MeV.
Our findings agree well with these estimations. 

Three-- and more--nucleon systems have been considered so far up to next--to--next--to--leading order 
\cite{Ep02}.  For the first time, the complete chiral three--nucleon force has been included in few--body
calculations, which starts to contribute at this order. N$^3$LO analysis of $>2N$ systems is in progress. 

Recently we have worked out the leading and subleading isospin--violating 3N forces using the method of 
unitary transformation \cite{EMP}, which are largely driven by nucleon and pion mass differences. 
The work on isospin--breaking 2N forces is underway.

I thank W.~Gl\"ockle, U.--G.~Mei{\ss}ner, H.~Kamada, A.~Nogga, J.E.~Palomar and H.~Wita{\l}a 
for enjoyable collaboration on these and other topics.

\begin{thebibliographynotitle}{12}
\bibitem{weinb} S.~Weinberg, Nucl. Phys.  B {\bf 363}  (1991) 3.
\bibitem{ord}  C.~Ord\'{o}\~{n}ez, L.~Ray and U.~van Kolck, 
              Phys. Rev. C {\bf 53} (1996) 2086.
\bibitem{EGM1} E.~Epelbaum, W.~Gl\"ockle and U.-G. Mei{\ss}ner,
  Nucl.\ Phys.\ A {\bf 637} (1998) 107.
\bibitem{EGMn3} E.~Epelbaum, W.~Gl\"ockle and U.-G. Mei{\ss}ner,
  Nucl.\ Phys.\ A {\bf 747} (2005) 362.
\bibitem{EM} D.~R.~Entem and R.~Machleidt, Phys.\ Rev.\ C {\bf 68} (2003) 041001.
\bibitem{Ep02}  E.~Epelbaum et al., Phys.\ Rev.\ C {\bf 66}, (2002) 064001.
\bibitem{EMP} E.~Epelbaum, U.~G.~Mei{\ss}ner and J.~E.~Palomar, 
to appear in Phys. Rev. C.
\end{thebibliographynotitle}

\newabstract 
\begin{center}
{\large\bf Charge-Symmetry-Breaking Nuclear Forces}\\[0.5cm]
{\bf U. van Kolck}\\[0.3cm]
Department of Physics, University of Arizona,\\
Tucson, AZ 85750, USA\\[0.3cm]
\end{center}

Charge symmetry is a particular isospin transformation.
While electromagnetic
interactions break isospin in general, the quark-mass difference breaks 
charge symmetry in particular. Charge-symmetry-breaking (CSB) quantities can
thus be linear in the quark-mass difference.
An example is the nucleon mass difference, whose
two main components are linear in the quark masses and in
the fine-structure constant, respectively.
These components can be separated by the pion interactions
they generate \cite{vK1}, which are different because of
the different ways quark masses and electromagnetism
break chiral symmetry explicitly.

An important source
of CSB is the nuclear potential, 
whose derivation is facilitated by a field redefinition
that eliminates the nucleon mass difference from asymptotic states
\cite{vK2}.
Isospin-violating effects
can be organized according to 
the expansion parameters associated with
the quark mass difference, $\epsilon  (m_\pi/m_\rho)^2$ 
(with $\epsilon  =(m_u-m_d)/(m_u+m_d)$), and electromagnetism,
$\alpha\sim \epsilon (m_\pi/m_\rho)^3$ (numerically) \cite{vK1}.
The leading CSB components of the two- and three-nucleon potentials
have been obtained in Refs. \citethree{vK3}{vK4}{vK2}
and in Refs. \citetwo{vK7}{vK5}, respectively.

Estimates \citetwo{vK4}{vK5} show that 
the $pp$-$nn$ scattering-length and 
$^3$He-$^3$H binding-energy differences
can be explained with natural parameters.
The best hope for a separation of nucleon-mass components 
probably rests on pion production \cite{vK6}.

\newabstract 
\begin{center}
{\large\bf Consistency of Weinberg's approach to the few-nucleon
problem
in EFT}\\[0.5cm]
D.~Djukanovic$^1$, M.~R.~Schindler$^1$, {\bf J. Gegelia}$^{1,2}$ and S.~Scherer$^1$\\[0.3cm]
$^1$Institut f\"ur Kernphysik, Johannes Gutenberg-Universit\"at,\\
D-55099 Mainz, Germany.\\[0.3cm]
$^2$High Energy Physics Institute, Tbilisi State University,
Tbilisi,
Georgia.\\[0.3cm]
\end{center}

   Weinberg's approach to the few-nucleon sector of EFT \cite{Weinberg:rz} has
encountered various problems.
   They originate from the renormalization of the LS equation with
non-renormalizable potentials. A consistent subtractive
renormalization requires the inclusion of an infinite number of
counterterm contributions and it has been argued that due to this
problem Weinberg's approach is inconsistent. To address the issue
of consistency and make some of the ''abstract arguments'' of
Ref.~\cite{Gegelia:2004pz} more explicit we have introduced a new
formulation of baryon chiral perturbation theory
\cite{Djukanovic:2004px}. While preserving all symmetries of the
effective theory, it leads to equations in the few nucleon (NN,
NNN, etc.) sector which are free of divergences and therefore one
does not need to include the contributions of an infinite number
of counterterms. The new formulation improves the ultraviolet
behavior of propagators and can be interpreted as a smooth cutoff
regularization scheme. Unlike the usual cutoff regularization, our
'cutoffs' are parameters of the Lagrangian and do not have to be
removed. Our new formulation is equivalent to the standard
approach and is equally well defined in the vacuum, one- and
few-nucleon sectors of the theory. It preserves all symmetries and
therefore satisfies the Ward identities.

    To improve the ultraviolet behavior of the propagators we introduce
additional {\it symmetry-preserving} terms into the Lagrangian.
These additional terms do not render all loop diagrams finite.
However, the remaining divergent diagrams contribute either in
physical quantities of the vacuum and the one-nucleon sectors, or
they appear as sub-diagrams in the potentials of the few-nucleon
sector. Therefore these diagrams can be regularized using standard
dimensional regularization and the iterations of the LS equation
do not generate any additional divergences.

\newabstract 
\begin{center}
{\large\bf Renormalization of the 1$\pi$ exchange interaction in higher partial waves}\\[0.5cm]
{\bf Andreas Nogga}$^1$\\[0.3cm]
$^1$Forschungszentrum J\"ulich, D-52425 J\"ulich, Germany.\\[0.3cm]
\end{center}

Chiral perturbation theory was successfully applied to the nucleon-nucleon system 
in the past \citethree{ordonez}{entem}{epelbaum}. The approach is based on the resummation 
of reducible diagrams using a Lippmann-Schwinger (LS) equation, which needs to be regularized
using a cutoff $\Lambda$.  But the singularity of the potentials entering the
LS equation necessarily generates a strong $\Lambda$ dependence in some partial waves.
In practice, this restricts $\Lambda$  to values well below the typical
 $\Lambda_{QCD} \approx 1$ GeV. 

This has been studied for the $S$-waves before \citetwo{beane}{arriola}.
Here, we study the cutoff dependence for the leading $1\pi$ exchange quantitatively also for 
the higher partial waves \cite{nogga}. 
We confirm a strong, periodic dependence on $\Lambda$ in some partial waves \cite{meissner} 
for the leading order 1$\pi$ exchange and 
show that this dependence can be absorbed into one counter term for each of these 
partial waves. The energy dependence of the resulting phase shifts is in good agreement 
with the data. 

Alternatively, we find that, for low energies, ranges of cutoffs exist for which the phase shifts 
only mildly depend  on $\Lambda$. The predictions for those cutoffs do also agree with the data.
Therefore, we propose two strategies.  For 
most higher partial waves, the ranges with mild cutoff dependence are large. 
For the leading $1\pi$ exchange interaction, $\Lambda \approx 8$ fm$^{-1}$ is a sensible choice 
and we find good agreement with the data in most cases.
For some lower partial waves, most prominently the $^3$P$_0$,  the regions of mild 
$\Lambda$ dependence are not large. Here, it seems 
advisable to promote counter terms, which can then absorb the $\Lambda$ dependence completely. 
In this case also the 3N binding energy is $\Lambda$ independent.

\newabstract 
\begin{center}
{\large\bf Renormalization Group Approach to NN-Scattering with Pion
       Exchanges: Removing the Cut-Offs~\footnote{Supported in
       part by funds provided by the Spanish DGI with grant
       no. BMF2002-03218, Junta de Andaluc\'{\i}a grant no. FM-225 and
       EURIDICE grant number HPRN-CT-2003-00311.}
}\\[0.5cm]
M. Pav\'on Valderrama and {\bf E. Ruiz Arriola}\\[0.3cm]
Departamento de F\'{\i}sica Moderna, \\ 
Universidad de Granada, E-18071 Granada,
       Spain. \\[0.3cm]
\end{center}

A non perturbative renormalization scheme for
Nucleon-Nucleon interaction based on boundary conditions at short
distances is presented and applied to the One and Two Pion Exchange
Potential. It is free of off-shell ambiguities and ultraviolet
divergences, provides finite results at any step of the calculation
and allows to remove the short distance cut-off in a suitable
way. Actually we see that our approach is equivalent to the Variable
S-matrix approach and offers a unique way to extract low energy
threshold parameters for a given NN potential. We extract those
parameters for the np system from the NijmII and Reid93 potentials, to
all partial waves with total angular momentum $j \le 5$. After having
done that, low energy constants and their non-perturbative evolution
can directly be obtained from experimental threshold parameters in a
completely unique and model independent way when the long range
explicit pion effects are eliminated. This allows to compute
scattering phase shifts which are, by construction consistent with the
effective range expansion to a given order in the C.M. momentum
$p$. In the singlet $^1S_0$ and triplet $^3S_1-^3D_1$ channels
ultraviolet fixed points and cycles are obtained respectively for the
threshold parameters, and consequently for the low energy
constants. This explains why it has been difficult to remove the
cut-offs by performing large scale fits to the data. We find that,
after properly removing the cut-off, scattering data are described
satisfactorily up to CM momenta of about $p \sim m_\pi$.

\newabstract 
\begin{center}
{\large\bf Chiral Perturbation Theory for Heavy Nuclei}\\[0.5cm]
{\bf Luca Girlanda}$^1$, Akaki Rusetsky$^{2,3}$ and Wolfram Weise$^4$\\[0.3cm]
$^1$ECT$^*$, Strada delle Tabarelle 286,\\
I-38050 Villazzano, Trento, Italy\\[0.3cm]
$^2$HISKP (Theorie), University of Bonn,\\
    Nu\ss{}allee 14-16, 53115 Bonn, Germany\\[0.3cm]
$^3$HEPI, Tbilisi State University,\\ University St.~9, 380086
    Tbilisi, Georgia\\[0.3cm]
$^4$Physik-Department, TU M\"{u}nchen, \\
D-85747 Garching, Germany\\[0.3cm]

\end{center}

We present an extension of the In-Medium Chiral Perturbation Theory
\cite{rhoChPT}, in   
which the nuclear background is characterized by a static, non-uniform
distribution of the baryon number describing the finite nucleus
\cite{grwpaper}. 
The nuclear structure information is encoded in a set of nuclear matrix
elements of free-nucleon field operators. The chiral counting applied to
such matrix elements allows to reduce considerably the nuclear input needed.
As an illustration the charged pion self-energy in the background of a
heavy nucleus is calculated at $O(p^5)$ of the chiral expansion and
the complete set of terms in the pion-nucleus optical potential arising
at this order is generated. We are able to identify  unambiguously
the nuclear finite size effects 
and disentangle the $S-$, $P-$ and $D-$wave contributions to the
optical potential without invoking the local density approximation. 
We include consistently the complete isospin violating effects arising
at order $O(p^5)$, including electromagnetic effects, which were not
taken into account in the literature.  Our analysis
only concerns the leading (linear) terms in density, because
these are the only ones showing up at this order. However it is well
known  that non-linear terms, coming from double
scattering diagrams or pion absorption, give important contributions
to the  binding energies and widths of pionic atoms. For these
reasons, in  order for our analysis to be relevant phenomenologically,
it should  be extended to chiral order $O(p^6)$. Whether our framework
can ben straightforwardly extended to this order, however, remains to
be seen.

\newabstract 
\begin{center}
{\large\bf Chiral Dynamics of Nuclear Matter: Role of Two-Pion\\ Exchange with 
Virtual Delta-Isobar Excitation}\\[0.5cm]
Stefan Fritsch, {\bf Norbert Kaiser} and Wolfram Weise\\[0.3cm]
Physik Dept. T39, TU M\"unchen, D-85747 Garching, Germany.\\[0.3cm]
\end{center}
We extend a recent three-loop calculation of nuclear matter \cite{nucmat} in 
chiral perturbation theory by including the effects from two-pion exchange
with single and double virtual $\Delta(1232)$-isobar excitation
\cite{deltamat}. Regularization dependent short-range contributions from
pion-loops etc. are encoded in a few NN-contact coupling constants. The 
empirical saturation point of nuclear matter, $\bar E_0 = -16\,$MeV, $\rho_0 
=0.16\,$fm$^{-3}$, can be well reproduced by adjusting the strength of a 
two-body term linear in density (and tuning an emerging three-body term 
quadratic in density). The nuclear matter compressibility comes out as $K= 
304\,$MeV. The real single-particle potential $U(p,k_{f0})$ 
\cite{pot} is substantially improved by the inclusion of the chiral $\pi N
\Delta$-dynamics: it grows now monotonically with the nucleon momentum $p$. 
The effective nucleon mass at the Fermi surface takes on a
realistic value of $M^*(k_{f0})=0.88M$. As a consequence, the critical
temperature of the liquid-gas phase transition \cite{liquidgas} gets lowered
to the value $T_c \simeq 15\,$MeV. We continue the complex
single-particle potential $U(p,k_f)+i\,W(p,k_f)$ also into the region above
the Fermi surface $p>k_f$. Furthermore, we find that the isospin properties of
nuclear matter get significantly improved by including the chiral $\pi
N\Delta$-dynamics. Instead of bending downward above $\rho_0$ as in previous
chiral calculations \cite{nucmat}, the energy per particle of pure neutron 
matter $\bar E_n(k_n)$ and the asymmetry energy $A(k_f)$ now grow 
monotonically with density \cite{deltamat}. In the density regime 
$\rho= 2\rho_n<0.2\,$fm$^{-3}$ relevant for conventional nuclear physics our
results agree well with sophisticated many-body calculations and 
(semi)-empirical values. Furthermore, we calculate the spin-asymmetry energy 
$S(k_f)$ \cite{spinstab} and find that the inclusion of the chiral $\pi N
\Delta$-dynamics is crucial in order to guarantee the spin-stability of 
nuclear matter: $S(k_f)>0$. Finally, the density dependent Landau parameters
$f_0(k_f),\, f_1(k_f),\,f_0'(k_f),\, g_0(k_f),\,g_0'(k_f),\, h_0(k_f),\,
h_0'(k_f)$ are calculated in the same framework. These quantities reveal the
spin- and isospin dependent interaction (including tensor components) of
quasi-nucleons at the Fermi surface.\\[-1cm]

\newabstract 
\begin{center}
{\large\bf Limit Cycle Physics}\\[0.5cm]
{\bf H.-W. Hammer}\\[0.3cm]
Institute for Nuclear Theory, University of Washington\\
Seattle, WA 98195-1550, USA \\[0.3cm]
\end{center}

The renormalization group (RG) is an important tool  
in many branches of physics. Its applications range from critical
phenomena in condensed matter physics
to the nonperturbative formulation of quantum field theories in
particle physics. While most RG flows show a simple fixed point 
behavior, more complex solutions are possible as well.
Wilson suggested already in 1971 that RG solutions could display
a limit cycle behavior \cite{Wil71}. 

One important example of a theory with a limit cycle
is the effective field theory (EFT) for 
non-relativistic three-body systems with large
scattering length \cite{BHK99}. Its applications range from
cold atoms to light nuclei. 
The large scattering length $a$ leads to universal properties
independent of the short-distance dynamics. In particular, one
can derive universal expressions for three-body observables with
log-periodic dependence on $a$ and the three-body parameter $\Lambda_*$. 
Furthermore, there are universal scaling functions relating different 
few-body observables. For a detailed review of these properties,
see Ref.~\cite{BrH04}.
                                                                              
The success of this EFT for nuclear few-body systems demonstrates 
that QCD is close to an infrared limit cycle. We have conjectured, that
QCD could be tuned to the critical trajectory for the limit cycle
by adjusting the up and down quark masses. The limit cycle would then
be manifest in the Efimov effect for the triton \cite{BrH03}.
It may be possible to demonstrate the existence of this
infrared RG limit cycle in QCD using Lattice QCD and EFT.
                                                                               
In two spatial dimensions, there is no limit cycle and no
three-body parameter $\Lambda_*$ because of the 
$c$-theorem. However, for bosons with a weakly bound
dimer, asymptotic freedom leads
to remarkable universal properties of $N$-boson droplets, such as
an exponential behavior of binding energies and droplet sizes 
\cite{Hammer:2004as}.

\newabstract 
\begin{center}
{\large\bf What have we learned so far about dilute fermions in the unitary limit? }\\[0.5cm]
Aurel Bulgac\\[0.3cm]
Department of Physics,
University of Washington\\
P.O. Box 351560, Seattle, WA 98195-1560, USA \\
\end{center}

In the last couple of years we have witnessed a tremendous progress in the field of
cold fermionic atoms, both experimentally and theoretically. After the first successful 
trapping of a fermionic atomic cloud \cite{jin} a  real breakthrough was the creation and the
subsequent study of the expansion of a strongly interacting degenerate
Fermi gas\cite{duke}. After that experimentalists have been able to
study the formation of extremely weakly bound molecules \cite{molecules}, the decay properties
of ensembles of such dimers\cite{decay}, the BEC of dimers\cite{bec},
a number of features of the BCS to BEC crossover\cite{bcsbec}, the
collective oscillations\cite{oscillations}, the formation of some kind
of condensate, with some still unclear properties\cite{condensation}
and finally the appearance of a gap in the excitation
spectrum\cite{gap}.

Eagles, Leggett and others have envisioned theoretically such a BCS to BEC
crossover\cite{leggett} and were able to describe qualitatively its
main features. Qualitative features of the BCS dilute atomic Fermi
superfluid have been discussed by a number of authors in recent
years\cite{eddy}. The theoretical description was based essentially on
the weak coupling BCS formalism, which is known to over predict the value
of the gap by a significant factor\cite{gorkov}. The crossover theory
of Leggett and its followers was based on a more or less
straightforward extension of the weak coupling BCS formalism to the
strong coupling regime. In the BEC limit there is an equally
significant correction of this results\cite{amm}. As it was noted by
Bertsch\cite{george}, a dilute Fermi system acquires universal
properties at, what nowadays we call, the Feshbach resonance. The
initial studies of the Bertsch MBX challenge showed that such a
system is stable\citetwo{baker}{heiselberg}. Only relatively recently that was
confirmed both theoretically\citethree{carlson}{chang}{giorgini} and
experimentally\cite{duke}.

While the race is still
on for providing the compelling evidence for superfluidity of such systems, many 
experimental results are in clear agreement with its existence \cite{abgfb1}. On the theoretical side
our overall understanding of these remarkable many body systems has improved tremendously, even 
though many questions remain yet unanswered. I shall present a review of the major experimental
results and of their present theoretical interpretation, along with a shopping list of issues awaiting
their resolution either in experiments and/or in theory.

\begin{thebibliographynotitle}{12}

\bibitem{jin} B. DeMarco and D.S. Jin, Science, {\bf 285}, 1703
(1999); K.M. O'Hara {\it et al.}, Phys. Rev. Lett. {\bf 82}, 4204 (1999).

\bibitem{duke} K.M. O'Hara, {\it et al.,} Science, {\bf 298}, 2179
(2002); M.E. Gehm, {\it et al.,} Phys. Rev. A {\bf 68}, 011401
(2003); T. Bourdel, {\it et al.,} Phys. Rev. Lett. {\bf 91}, 020402
(2003).

\bibitem{molecules} C. A. Regal, {\it et al.,} Nature {\bf 424}, 47
(2003); K.E. Strecker, {\it et al.,} Phys. Rev. Lett. {\bf 91},
080406 (2003); J. Cubizolles, {\it et al.,} Phys. Rev. Lett. {\bf
91}, 240401 (2003); S. Jochim, {\it et al.,} Phys. Rev. Lett. {\bf
91}, 240402 (2003).

\bibitem{decay} K. Dieckmann, {\it et al.,} Phys. Rev. Lett. {\bf
  89}, 203201 (2002); C.A. Regal, {\it et al.,}
  Phys. Rev. Lett. {\bf 92}, 083201 (2004); see also S. Jochim, {\it
    et al.,} in Ref. \cite{molecules}.

\bibitem{bec} M. Greiner, {\it et al.,} Nature {\bf 426}, 537 (2003);
M.W. Zwierlein, {\it et al.,} Phys. Rev. Lett. {\bf 91}, 250401
(2003); S. Jochim, {\it et al.,} Science {\bf 302}, 2101 (2003).

\bibitem{bcsbec} M. Bartenstein, {\it et al.,} Phys. Rev. Lett. {\bf
92}, 120401 (2004); T. Bourdel {\it et al.,} cond-mat/0403091.

\bibitem{oscillations} J. Kinast {\it et al.,}
Phys. Rev. Lett. {\bf 92}, 150402 (2004); M. Bartenstein, {\it et
al.,} Phys. Rev. Lett. {\bf 92}, 203201(2004).

\bibitem{condensation} C.A. Regal, {\it et al.,}
  Phys. Rev. Lett. {\bf 92}, 040403 (2004); M.W. Zwierlein, {\it et
  al.,} Phys. Rev. Lett. {\bf 92}, 120403 (2004).

\bibitem{gap} C. Cheng, {\it et al.,} Science, {\bf 305}, 1128 (2004).

\bibitem{leggett} D.M. Eagles, Phys. Rev. {\bf 186}, 456 (1969);
A.J. Leggett, in {\it Modern Trends in the Theory of
Condensed Matter}, eds. A. Pekalski and R. Przystawa,
Springer--Verlag, Berlin, 1980; J. Phys. (Paris) Colloq. {\bf 41},
C7--19 (1980); P. Nozi\`eres and S. Schmitt--Rink, J. Low
Temp. Phys. {\bf 59}, 195 (1985); C.A.R. S\'a de Mello {\it et al.,}
Phys. Rev. Lett. {\bf 71}, 3202 (1993); M. Randeria, in {\it
Bose--Einstein Condensation}, eds. A. Griffin, D.W. Snoke and
S. Stringari, Cambridge University Press (1995), pp 355--392.

\bibitem{eddy} E. Timmermans, {\it et al.,} Phys. Lett. A {\bf 285},
228 (2001); M. Holland, {\it et al.,} Phys. Rev. Lett. {\bf 87},
120406 (2001); Y. Ohashi and A. Griffin, Phys. Rev. Lett. {\bf 89},
130402 (2002) and references therein.

\bibitem{gorkov} L.P. Gorkov and T.K. Melik--Barkhudarov,
Sov. Phys. JETP {\bf 13}, 1018 (1961); H.  Heiselberg, {\it et al.,}
Phys. Rev. Lett. {\bf 85}, 2418 (2000).

\bibitem{amm} P. Pieri and G.C. Strinati, Phys. Rev. B {\bf 61}, 15370
(2000); D.S. Petrov, {\it et al.,} Phys. Rev. Lett. {\bf 93}, 090404 (2004); A. Bulgac, {\it
et al.,} cond-mat/0306302.

\bibitem{george} G.F. Bertsch, {\it Many-Body X challange problem}, see
R.A. Bishop, Int. J. Mod. Phys. {\bf B 15}, {\it iii}, (2001).

\bibitem{baker} G.A. Baker, Jr., Int. J. Mod. Phys. {\bf B 15}, 1314
  (2001). 

\bibitem{heiselberg} H. Heiselberg, Phys. Rev. A {\bf 63}, 043606 (2001).

\bibitem{carlson} J. Carlson, {\it et al.,} Phys. Rev. Lett. {\bf
  91}, 050401 (2003).

\bibitem{chang} S.Y. Chang, {\it et al.,} Phys. Rev. A {\bf 70}, 043602 (2004).

\bibitem{giorgini} G.E. Astrakharchik, {\it et al.,} cond-mat/0406113, Phys. Rev. Lett., in press.

\bibitem{abgfb1} A. Bulgac and G.F. Bertsch, cond-mat/0404301; {\it ibid}
A. Bulgac and G.F. Bertsch, cond-mat/0404687, Phys. Rev. Lett. {\bf 94}, in press (2005).

\end{thebibliographynotitle}

\newabstract 
\begin{center}
{\large\bf An Effective Theory for the Four-Body System}\\[0.5cm]
Lucas Platter\\[0.3cm]
HISKP (Abt. Theorie), Universit\"at Bonn,\\
Nussallee 14-16, 53115 Bonn, Germany\\[0.3cm]
Forschungszentrum J\"ulich, IKP(Th)\\
52425 J\"ulich, Germany.\\[0.3cm]
\end{center}
We use an effective theory with contact interactions to compute universal
 properties of four-body systems with a large two-body scattering length $a$.
Along with the corresponding power counting this effective theory is a
systematic expansion in $\ell/a$, where $\ell$ denotes the typical
low-energy scale of the underlying interaction. By generating
the leading order effective potential and employing the Yakubovsky equations
we are able to compute binding energies of four-body systems.
This approach has been applied to bosons \cite{Platter:2004qn}
(the $^4$He tetramer) and to fermions 
(the $\alpha$-particle) \cite{Platter:2004zs}.\\
A particular characteristic of this approach is, that in the three-body
system a three-body force at leading order is needed to renormalize
observables \cite{3boson}. It is not a priori clear whether a four-body
force is needed in the four-body system at leading order. However, 
an analysis of the renormalization properties of the four-body
system shows that this is not the case.
Further, it turns out that a well-known linear correlation between the
three-body binding energies and the four-body binding energies shows up 
naturally within this approach and can be considered as a consequence
of the large scattering length in the two-body subsystem.\\
In the future other observables like scattering amplitudes or electromagnetic
properties in the four-body sector will be computed.

\newabstract 
\begin{center}
{\large\bf Subtleties in pion production reactions on few nucleon 
systems}\\[0.5cm]
Christoph Hanhart\\[0.3cm]
Institut f\"ur Kernphysik,\\
Forschungszentrum J\"ulich,\\
52428 J\"ulich,\\
Germany\\[0.3cm]
\end{center}

Meson reactions on light nuclei were discussed. The focus was on 
the role of the nucleon recoil corrections in low energy meson--nucleus
interactions. We demonstrated explicitly when calculations within
the static approximation are justified and when the recoils need to be kept
explictly in the propagators, depending on whether the intermediate two
nucleon state is Pauli blocked or not, while the meson is in flight.

In reactions where a two nucleon intermediate state, that occurs while the
pion is in flight, is allowed the two nucleons may interact. We demonstrated
that---as a consequence of the large $NN$ scattering lengths---these
contributions are numerically significant and argue that this observation is
in line with the ideas of Weingbergs power counting.
 
As examples the reactions $\pi d\to \pi d$ and $\gamma d\to \pi^+ nn$ are
presented \cite{inprep}. Consequences for the chiral counting for reactions on
nuclei were discussed.

\newabstract 
\begin{center}
{\large\bf Neutral pion electroproduction off the deuteron}\\[0.5cm]
V.~Bernard$^1$, {\bf H.~Krebs}$^2$, Ulf-G.~Mei{\ss}ner$^{2,3}$\\[0.3cm]
$^{1}$Laboratoire de Physique Th\'eorique,
Universit\'e Louis Pasteur,\\
F-67037 Strasbourg Cedex 2, France,\\[0.3cm]
$^{2}$Helmholtz Institut f\"ur Strahlen- und Kernphysik (Theorie),
Universit\"at Bonn,\\ Nu\ss allee 14-16, D-53115 Bonn, Germany\\[0.3cm]
$^{3}$Institut f\"ur Kernphysik (Theorie), Forschungszentrum J\"ulich,\\
D-52425 J\"ulich, Germany\\[0.3cm]
\end{center}

I present our latest calculations on neutral pion electroproduction off deuterium in chiral perturbation theory. 
The interaction kernel and the deuteron wave functions are described
consistently with each other in a novel formalism of chiral nuclear EFT~\cite{KBMOrtho} based on the
$\hat{Q}$-box approach
of Kuo and collaborators~\cite{Suzuki1}. The interaction kernel decomposes into single
nucleon (impulse approximation) and three-body (meson exchange) pieces. 
Calculating the latter to third order in the chiral expansion~\cite{KBMq3} leads to a
satisfactory description of the data at photon virtuality $Q^2 = 0.1\,$GeV$^2$
from MAMI-B~\cite{Ewald}. The theoretical uncertainties at this chiral order 
given by the two different fit procedures we are using to fix the two low--energy constants related to the elementary pion electroproduction on a neutron amplitude are
of the same size as the error bars of the MAMI data.

The complete fourth order calculation of the 
three--body contribution~\cite{KBMq4}, with all new parameters fixed from the earlier studies of $\pi N$ and $NN$ phase shifts~\cite{EpelbImpr2}, 
leads to a decrease of the theoretical uncertainties and to a further improvement in the description of the data.

\newabstract 
\newcommand{\3}{{\ss}}

\newcommand{\MeV}{\mathrm{MeV}}
\newcommand{\EPJA}{\textit{Eur.\ Phys.\ J.\ }\textbf{A}}
\newcommand{\FBSS}{\textit{Few-Body Syst.\ }\textbf{Suppl.}}
\newcommand{\IJMPE}{\textit{Int.\ J.\ Mod.\ Phys.\ }\textbf{E}}
\newcommand{\JMathP}{\textit{J.\ Math.\ Phys.\ }}
\newcommand{\NP}{\textit{Nucl.\ Phys.\ }}
\newcommand{\NPA}{\textit{Nucl.\ Phys.\ }\textbf{A}}
\newcommand{\NPB}{\textit{Nucl.\ Phys.\ }\textbf{B}}
\newcommand{\PLB}{\textit{Phys.\ Lett.\ }\textbf{B}}
\newcommand{\PR}{\textit{Phys.\ Rev.\ }}
\newcommand{\PRep}{\textit{Phys.\ Rept.\ }}
\newcommand{\Phy}{\textit{Physica}\ }
\newcommand{\PRB}{\PR\textbf{B}}
\newcommand{\PRC}{\PR\textbf{C}}
\newcommand{\PRD}{\PR\textbf{D}}
\newcommand{\PRL}{\PR\textit{Lett.\ }}

\begin{center}
  {\large\bf Nucleon Polarisabilities from Elastic Compton Scattering}\\[0.5cm]
  Harald W.~Grie{\ss}hammer$^{1}
  $\\[0.3cm]
  $^1$Institut f{\"u}r Theoretische Physik (T39), TU M{\"u}nchen,
  Germany\\[0.3cm]
\end{center}

Chiral EFT
is the tool to accurately determine the proton and neutron spin-independent
and spin-dependent dipole polarisabilities. Dynamical polarisabilities,
defined by a multipole-analysis at fixed photon
energy~\citetwo{Griesshammer:2001uw}{polas2}, allow one to quantitatively
understand the dispersive effects from the nucleon's internal degrees of
freedom: Whereas pions suffice to describe data at less than $70$
MeV~\cite{beane}, the energy- and angular dependence
of the 
polarisabilities induced by the explicit $\Delta(1232)$-degree of freedom is
mandatory
at higher energies, 
in particular
to understand deuteron Compton scattering data at $95$
MeV~\cite{polasfromdeuteron}.

Predicting this energy-dependence at LO by Chiral Perturbation Theory with
explicit $\Delta$, one finds from all proton Compton data below $180\;\MeV$
the static values
$\bar{\alpha}^p=11.0\pm1.4_\mathrm{stat}\pm0.4_\mathrm{Baldin},\;
\bar{\beta}^p=2.8\mp1.4_\mathrm{stat}\pm0.4_\mathrm{Baldin}$ (in
$10^{-4}\;\mathrm{fm}^3$), which compares favourably both in magnitude and
error with alternative extractions~\cite{polas2}. For the static iso-scalar
dipole polarisabilities, deuteron Compton data at $69$ and $94\;\MeV$ gives
$\bar{\alpha}^s=
12.6\pm0.8_\mathrm{stat}\pm0.7_\mathrm{wavefu}\pm0.6_\mathrm{Baldin}$,
$\bar{\beta}^s=
1.9\mp0.8_\mathrm{stat}\mp0.7_\mathrm{wavefu}\pm0.6_\mathrm{Baldin}
$~\cite{polasfromdeuteron}.  Thus, proton and neutron polarisabilities are
identical within the accuracy of available data.  Both times, the Baldin sum
rule is well-matched in a free fit. We propose also to dis-entangle the thus
far ill-determined spin-dependent polarisabilities from asymmetries in
experiments with polarised beams and targets around the pion production
threshold both on the proton and neutron, and present
predictions~\cite{polas3}. Future precision experiments will allow for a
multipole-analysis, in which the energy-dependence of the dipole
polarisabilities can be probed directly even in the resonance region.

\newabstract 
\begin{center}
{\large\bf Effective Theory of Higgs-less Electroweak Symmetry Breaking}\\[0.5cm]
Jan Stern\\[0.3cm]
Groupe de Physique Theorique, IPN, \\
Universite de Paris-Sud, 91405 Orsay, France\\[0.3cm]
\end{center}

The generic low-energy effective theory (LEET) of Higgs-less electroweak
symmetry breaking is considered operating with naturally light gauge bosons
and chiral fermions in addition to three Goldstone bosons arising from a
symmetry breaking of $SU(2)_L \times SU(2)_R$. In such a theory, low-energy
and loop expansions are related by the same Weinberg power counting formula
as in the $ChPT$. Our goal is a systematic LEET ($p\ll 4\pi v \sim 3 TeV$),
in which symmetries and chiral power counting guarantee that at leading order
the theory coincides with the Standard Model without a Higgs particle 
Previous attempts failed to satisfy the latter requirement.
In order to suppress the $0(p^2)$ non-standard couplings, the LEET has to be
based on a symmetry $S_{nat}$ that is larger than $SU(2)_L\times U(1)_Y$,
including, in particular, the custodial symmetry. Exact $S_{nat}$ forbids
couplings of the three Goldstone bosons to elementary massless
$SU(2)\times SU(2) \times U(1)_{B-L}$ gauge fields and to fermions. 
The coupling
is introduced and the symmetry is reduced to the electroweak
$SU(2)\times U(1)_Y$ via constraints which eliminate the redundant fields
keeping track of the whole local symmetry $S_{nat}$. This procedure is
{\bf equivalent} to introducing non propagating scalar `spurion fields'' with
definite transformation properties under $S_{nat}$ and with vanishing
covariant derivatives \cite{HS1}. Spurions play a similar role as quark
masses  in
$ChPT$ describing a hierarchy of symmetry breaking effects.
The whole LEET is defined as a double expansion in powers of momenta and
spurions. At the leading order ($0(p^2)$ , no spurion insertion ) one
recovers  the
SM couplings of massive W and Z to massless fermion doublets with no scalar
particle left in the spectrum \cite{HS2}. The fermion masses are the first
manifestation  of spurions.
Further necessary consequence is a tiny lepton number violation and Majorana
neutrino masses. The NLO  ($0(p^2)$ , two spurion insertions) consists of
specific universal corrections to the vector boson - fermion vertices.
Power counting predicts the latter to contribute before loops and to be
potentially more important than the oblique corrections which first appear
at the NNLO.

\newabstract 
\begin{center}
{\large\bf From mooses to 5D and back to large-$N_c$ QCD ?}\\[0.5cm]
Johannes Hirn\\[0.3cm]

IFIC, Universitat de Val\`encia, Val\`encia, Spain\\[0.3cm]

\end{center}

Consider $K+1$ Goldstone boson (GB) multiplets, each describing the spontaneous breaking
$\mathrm{SU} \left( N_f \right)_{L_k} \times \mathrm{SU} \left( N_f \right)_{R_k} \to \mathrm{SU} \left( N_f \right)_{L_k
+ R_k}$, ($k = 0, \cdots, K$). Introduce $K$ $\mathrm{SU}\left(N_f \right)_{G_k}$ Yang-Mills theories. To obtain the open linear moose, identify each gauge group  $\mathrm{SU} \left(
N_f \right)_{G_k}$ ($k = 1, \cdots, K$), with $\mathrm{SU}
\left( N_f \right)_{R_{k - 1} + L_k}$. All vectors get masses via $K$ Higgs mechanisms, leaving one GB multiplet in the spectrum.

Sticking to tree level \cite{Kunszt:2004ps}, one can impose an {\em approximate locality} on the moose interpreted as a deconstructed fifth dimension. For $N_f = 2$, the model reproduces some interesting results involving pions and vector resonances~{\cite{Son:2003et}}.
However, the approximately local moose yields (at tree-level) $K$ generalized Weinberg sum rules (GWSRs) {\cite{Hirn:2004ze}}, as opposed to two in  QCD. Note tat the GWSRs are not related to a 
symmetry of the moose lagrangian, but reflect the approximate
locality.

Nevertheless, the GWSRs can be interpreted as a consequence of a
larger symmetry reduced by spurions \cite{Hirn:2004ze}. In this way, one gains a perturbative 
control of the fifth dimension locality and corrections to the GWSRs
in a systematic low-energy expansion.
Still, in this expansion, the masses of the resonances are
expansion parameters $m \ll 4 \pi f_{\pi}$, which is unrealistic for QCD, except for large $N_c$~{\cite{Contino:2003ve}}: the moose provides a general framework for the introduction of
{\em light} massive vector fields into a low-energy effective theory of Goldstone bosons.

Extrapolating to a five-dimensional {\em model}, one expects an
infinite set of GWSRs {\cite{Barbieri:2003pr}}. In a Randall-Sundrum set-up, we propose a model with two $\mathrm{U} \left( N_f \right)$
bulk gauge fields (identified on the IR brane) and bulk scalars (to account for a non-zero quark condensate by providing a shortcut between the left and right chiral sources, as well as to introduce (pseudo)-scalar resonances) \cite{everybody}.
We {\em ask} whether this hadronic model can reproduce high-energy perturbative QCD results (two Weinberg sum rules, $\Pi_{V V}
\left( Q^2 \right) \sim \ln Q^2$~{\cite{Son:2003et}}, Brodsky-Lepage behavior).

\newabstract 
\begin{center}
{\large\bf Charm-strange mesons}\\[0.5cm]
{\bf Felix Sassen}$^1$, S. Krewald$^1$, U.-G. Mei\ss ner$^{1,2}$ and C. Hanhart$^1$\\[0.2cm]
$^1$FZ J\"ulich, IKP(Th), D-52425 J\"ulich, Germany.\\
$^2$Univ. Bonn, HISKP(Th), D-53115 Bonn, Germany.\\[0.3cm]
\end{center}

The charm-strange resonances $D^*_{sJ}(2317)$, $D_{sJ}(2460)$ and
$D_{sJ}(2632)$ discovered by the BaBar, CLEO and Selex collaborations may be
accommodated in
the spectrum expected for a heavy-light system by the chiral doubler
scenario\cite{Nowak}. Meson-meson-dynamics is considered as a
possible mechanism to generate   the $D^*_{sJ}(2317)$ \cite{Barnes}.
 We present an analysis
of the interaction in the $KD$-channel to check this
proposition. As one would have expected from the quark structure, which is very
similar to the $K\bar{K}$ channel, an isoscalar molecule is formed in the
$KD$ channel. Therefore the question arises how to discriminate a chiral doubler from
a molecule. One way to distinguish a molecular state from a genuine state was
suggested long ago by Weinberg with respect to the structure of the
deuteron\cite{Weinberg}.
 He introduced the normalization factor
$Z=\sum_i\left|\left<D_{s0}^{*i}|D_{sJ}^{*}(2317)\right>\right|$ to define the
elementary content ($\sum_i D_{s0}^{*i}$)  of an observed state
($D_{sJ}^{*}(2317)$) and related it to the effective range parameters $a$ and
$r$ via
$$ a=-\frac{2(1-Z)}{2-Z}R+\mathcal{O}\left(\frac{1}{\beta}\right)\qquad\textrm{
  and }\qquad r=-\frac{Z}{1-Z}R+\mathcal{O}\left(\frac{1}{\beta}\right).
$$
Here $1/\beta$ denotes the range of the binding force and $R=1/\sqrt{2\mu\epsilon}$
depends on the binding energy $\epsilon$ and the reduced mass $\mu$. We find this
approach to work even in the inelastic case, as indicated in ref.\cite{Baru}.
However it may break down if two  poles are close to the
threshold.
 The  molecular 
picture predicts a larger width
for the $D_{sJ}^{*}(2317)$ than the chiral doubler scenario does, if isospin
violations via $\pi\eta$-mixing and via mass differences of charged and
neutral mesons are considered.  
We checked our $SU(4)$
estimates for the coupling constants by calculating the decay widths of the
$D^{*+}$ and comparing the results to experimental data. The agreement was rather 
good and changing the couplings within reasonable bounds  does not alter 
our results.

\newabstract 
\begin{center}
{\large\bf Nuclear Physics from Lattice QCD
}\\[0.5cm]
Martin J. Savage\\[0.3cm]
Department of Physics, University of Washington\\
Seattle, Washington, USA
\end{center}

Determination of the properties and interactions of the deuteron directly 
from QCD will establish  a  milestone in theoretical physics.
As lattice QCD is the only known method for rigorous QCD calculations, 
it will be a coherent effort in lattice QCD and other areas of 
theoretical nuclear physics that  first achieves this goal.
As the deuteron is the simplest nucleus, such computations will
usher in a new era of nuclear physics, an era in which model-independent 
calculations of nuclei and their interactions will, for the first-time,
become possible.
A significant effort is already underway toward this goal.
In addition to LHPC~\footnote{
The Lattice Hadron Physics Collaboration.
http://www.jlab.org/$\sim$dgr/lhpc/ .
},
one lattice collaboration,
NPLQCD~\footnote{The NPLQCD collaboration.
http://www-nsdth.lbl.gov/$\%$7Ebedaque/nplqcd/nplqcd\_frame.html .},
has been formed within the last year whose main
objective is to perform such calculations
which are only now becoming feasible due to advances in technology and
in theoretical nuclear physics.

Within the last couple of years, a number of papers have explored
the requirements for and constraints on 
rigorously extracting the properties of the two-nucleon 
sector, the NN phases-shifts, electroweak matrix elements and static 
properties of the deuteron~\citetwo{Beane:2003da}{Detmold:2004qn}.
While it will be comforting to recover these well-known quantities, 
it is just a first step to computing interactions in regions that are not 
accessible to experiment.
This analysis has been extended to systems containing strange quarks, where
the theoretical framework for extracting hyperon-nucleon interactions 
from lattice QCD has been put in place~\cite{Beane:2003yx}.
As of January 2005, the NPLQCD collaboration was awarded 
$7\times 10^5$ processor-hours on the JLab cluster by SciDAC, for studies 
that include the $NN$, the $\Lambda N$ and $\Lambda\Lambda$ 
interactions on the publicly available dynamical, staggered MILC lattices.

\newabstract 
\begin{center}
{\large\bf Nucleons and nuclei from lattice QCD}\\[0.3cm]
{\bf Silas R.~Beane}$^{1,2}$\\[0.1cm]
$^1$Dept. of Physics, University of New Hampshire,
Durham, NH 03824-3568.\\
$^2$Jefferson Laboratory, 12000 Jefferson Avenue,
Newport News, VA 23606.
\end{center}

\noindent Despite remarkable technical advances, lattice QCD
simulations necessitate the use of quark masses, $m_q$, that are
significantly larger than the physical values, lattice spacings, $a$,
that are not significantly smaller than the physical scales of
interest, and lattice sizes, $L$, that are not significantly larger
than the pion Compton wavelength. Fortunately, in many cases, the
dependence of hadronic physics on these parameters can be calculated
analytically in the low-energy effective field theory, thus
allowing rigorous extrapolations to remove lattice
artifacts. Sometimes lattice artifacts are critical to the
extraction of physics from the simulation.  The Maiani-Testa
theorem~\cite{Maiani:ca} precludes determination of scattering
amplitudes away from kinematic thresholds from Euclidean-space Green
functions at infinite volume. By generalizing a result from
non-relativistic quantum mechanics to quantum field theory,
L{\"u}scher~\cite{Luscher:1990ux} realized that one can access
$2\rightarrow 2$ scattering amplitudes from lattice simulations
performed at {\it finite} volume. This method opens up the
study of nuclear physics to lattice QCD simulations.  (Only one
lattice QCD calculation of the nucleon-nucleon scattering
lengths~\cite{Fukugita:1994ve} has been attempted.)  There is a
sizable separation of length scales in nuclear physics, and as a
result, the scattering lengths in both $S-$wave channels are
unnaturally-large compared to all typical strong-interaction length
scales, including the range of the nuclear potential, which is
determined by the pion Compton wavelength. Perhaps
counter-intuitively, in simulating two-nucleon processes, the relevant
lengths scales are those of the nuclear potential and {\it not} the
scattering lengths, and thus as long as the lattice is large compared
to the inverse of the pion mass one can in principle ``simply ''
determine matrix elements and scattering parameters~\cite{Beane:2003da}. Recently a
new lattice QCD collaboration, NPLQCD~\cite{nplqcd}, has formed to begin lattice
QCD simulations of simple nuclear systems.

\newabstract 

\begin{center}
{\large\bf Towards a lattice determination of chiral NLO coefficients}\\[0.5cm]
{Stephan D\"urr}$^1$\\[0.3cm]
$^1${Universit\"at Bern, ITP, Sidlerstr.\,5, 3012 Bern, Switzerland}\\[0.3cm]
\end{center}


For those QCD low-energy constants that describe the quark mass dependence of
Green's functions, e.g.\ $l_3,l_4,l_7$ in the $SU(2)$ framework of XPT
\cite{Gasser:1983yg}, a lattice determination seems promising, since there $m$
is a parameter which can, in principle, take arbitrary values.
In practice three (major) complications arise.

\noindent
\begin{minipage}{\textwidth}
\vspace{1mm}
\begin{minipage}{7.1cm}
\includegraphics[width=7cm]{durrfig.eps}
\end{minipage}
\begin{minipage}{7.2cm}
\hspace{5mm}
First, it gets expensive to take $m$ light, both in the
quark propagators and in the functional determinant.
Second, the continuum limit must be taken, since cut-off effects may be
particularly large close to the chiral limit.
That $ma\!\ll\!1$ (and $Ma\!\ll\!1$) is not sufficient to have small
discretization errors is borne out in the figure to the left, see
\cite{Durr:2004ta}.
Here the scalar condensate in the Schwinger model is plotted versus the
quark mass, and only one of the two formulations re-
\end{minipage}
\vspace{0.5mm}
\end{minipage}
produces the analytic result
$\lim_{m=0}\langle\bar\psi\psi\rangle/e\!=\!0.1599...$ within errors.
Thus, cut-off effects can completely mask the underlying continuum physics.
The third complication is that even in todays dynamical ($N_{\!f}\!=\!2$)
simulations the way the scale is set still matters.
I have compared the NLO chiral prediction for the degenerate
($m\!=\!m_u\!=\!m_d$) quark mass dependence of the pseudo-Goldstone boson mass
to the perturbatively renormalized data (at 1-loop) 
\citetwo{AliKhan:2001tx}{Durr}.
Setting the scale through $r_0$ the data are consistent with the chiral
prediction, even if one fixes the parameter $F_\pi$ to its physical value.
Notably, there is no indication that $l_3$ could be different from the original
GL estimate \cite{Gasser:1983yg}, but it is hard to turn this into a positive
statement, since only some of the data are likely in a regime where XPT
applies \cite{Durr}.
Furthermore, regarding the consistency with XPT Aoki reaches an adverse
conclusion, based on the same data, after setting the scale through
$M_\rho$ \cite{Aoki:2003yv}.
Thus, the only safe prediction is that the issue will be with us for some
time.

\newabstract 
\begin{center}
{\large\bf   Going chiral: overlap and twisted mass fermions}\\[0.5cm]
Karl Jansen$^1$\\[0.3cm]
$^1$John von Neumann-Institute for Computing,\\
Platanenallee 6, 15738 Zeuthen, Germany\\[0.3cm]
\end{center}

We present lattice simulations that aim at reaching small values of 
the pion mass
such that contact to chiral perturbation theory can safely be made. 
To this end we compare Wilson twisted mass and chirally invariant 
overlap fermions and found 
that both formulations of lattice QCD can reach this goal \cite{ref1} since 
with them
quenched simulations at $m_\pi\approx 230$MeV become possible.
However, Wilson twisted 
mass fermions cost a factor 10-40 less computertime than overlap fermions
\cite{ref2}. In refs.~\citetwo{ref3}{ref4} we used twisted mass fermions 
for studying 
the phase diagram of lattice QCD and found a surprising phase structure with 
pronounced signals of first order phase transitions. 
We demonstrated in refs.~\citetwo{ref5}{ref6} that finite size effects may be the 
largest systematic error in the computation of moments of parton distribution 
functions on the lattice. Here it would be very helpful to have analytical
calculations, as e.g. from chiral perturbation theory, to describe these 
finite size effects.

\newabstract 

\def\mres{m_{\rm res}}
\def\leqx{\,\raisebox{-1.0ex}{$\stackrel{\textstyle <}{\sim}$}\,}

\begin{center}
{\large\bf Applications of ChPT to QCD with domain-wall fermions}\\[0.5cm]
{\bf Maarten Golterman}$^1$ and Yigal Shamir$^2$\\[0.3cm]
$^1$Department of Physics and Astronomy,
San Francisco State University\\
San Francisco, CA 94132, USA\\[0.3cm]
$^2$School of Physics and Astronomy,\\
Tel-Aviv University, Ramat~Aviv,~69978~ISRAEL.\\[0.3cm]
\end{center}

Reporting on recent work \cite{mgys},
we discuss the very different roles of the valence-quark and the sea-quark
residual masses ($\mres^v$ and $\mres^s$) in dynamical domain-wall fermions
simulations \cite{rbcdyn}.  Focusing on matrix elements
of the effective weak hamiltonian containing a power divergence
\cite{rbc},
we find that $\mres^v$ can be a source of a much bigger systematic error.
To keep all systematic errors due to residual masses
at the 1\% level, we estimate that one needs $a\mres^s \leqx 10^{-3}$
and  $a\mres^v \leqx 10^{-5}$, at a lattice spacing $a\sim 0.1$~fm,
if only the single power-divergent subtraction already present in the
continuum theory is performed.

\newabstract 
\begin{center}
{\large\bf Excited Hadrons from Lattice Calculations:\\ Approaching the Chiral Limit}\\[0.5cm]
Dirk Br\"ommel$^1$,
Tommy Burch$^1$, 
Christof Gattringer$^1$, 
Leonid Ya.\ Glozman$^2$, 
Christian Hagen$^1$, 
Dieter Hierl$^1$, 
Reinhard Kleindl$^2$,\\
{\bf C.\ B.\ Lang}$^2$, 
and Andreas Sch\"afer$^1$\\
(BGR [Bern-Graz-Regensburg] Collaboration) \\[0.3cm]
$^1$Institut f\"ur  Theoretische Physik, Universit\"at Regensburg,\\
D-93040 Regensburg, Germany\\[0.3cm]
Institut f\"ur Physik, FB Theoretische Physik, Universit\"at Graz,\\
A-8010 Graz, Austria\\[0.3cm]
\end{center}

Ground state spectroscopy for quenched lattice QCD is well understood. However,
it is still a challenge to obtain results for excited hadron  states
\cite{broemmeletal}. In our study we present results from a new approach for
determining spatially optimized operators for lattice spectroscopy of excited
hadrons. 

In order to be able to approach physical quark masses we work with the chirally
improved Dirac operator \cite{chirimp}, i.e.,  approximate Ginsparg-Wilson
fermions. Since these are  computationally expensive we restrict ourselves to a
few quark sources.  We use Jacobi  smeared quark sources with different widths
and combine them  to  construct hadron operators with different spatial wave
functions \cite{prd-paper}.  The cross-correlation matrix is then analyzed with
the variational method. This leads to optimized combinations of hadron
operators that provide us with better signals for the excited states.

This approach allows us to identify the Roper state and other excited baryons
and mesons, also in the  strange sector. We find that excited states may be
more affected by finite volume effects, but also be more sensitive to the 
quenched approximation. Finite volume studies,  including also scaling
properties, are under way.


\newabstract 
\begin{center}
{\large\bf The $\epsilon$-regime of QCD and its applications to
  non-leptonic Kaon decays}\\[0.2cm]
Hartmut Wittig\\
DESY, Notkestr. 85, 22603 Hamburg, Germany
\end{center}

A quantitative understanding of $K\to\pi\pi$ decays and the associated
$\Delta{I}=1/2$ rule has been elusive for many years. A successful
treatment in the context of QCD should reproduce the large ratio of
decay amplitudes, $A_0/A_2\approx22$, where the subscript labels the
isospin of the pion pair in the final state.
In ref.\,\cite{Giusti:2004an} we proposed a strategy which seeks to
disentangle various possible origins of the $\Delta{I}=1/2$ rule and
clarifies the specific r\^ole of the charm quark. The goal is a
precise determination of the LECs $g_1^{\pm}$, which appear in the
effective low-energy description of $\Delta{S}=1$ transitions, using
lattice simulations. These LECs are linked to the ratio $A_0/A_2$.
The r\^ole of charm can be studied by comparing the values of
$g_1^\pm$ obtained for an unphysically light charm quark $m_u=m_c$ to
those computed for $m_c\gg m_u$. Thus, in contrast to other lattice
studies, we keep the charm quark active. The other key ingredients of
our strategy are the use of overlap fermions, which preserve chiral
symmetry at non-zero lattice spacing, and a matching of QCD to ChPT
via the so-called $\epsilon$-regime. Chiral symmetry ensures that the
matching of lattice data to ChPT is on a solid footing. Furthermore,
it is guaranteed that the subtraction of power divergences can be
avoided at all stages of the calculation, provided that the charm
quark is active. The chiral counting rules of the $\epsilon$-regime
imply that no additional coupling terms with unknown coefficients are
generated at NLO in the treatment of $\Delta{S}=1$ transitions. Thus,
the matching of correlation functions of four-quark operators to the
expressions of ChPT can be easily performed at NLO.
The use of overlap fermions requires efficient numerical techniques
\citetwo{Giusti:2002sm}{Giusti:2004yp}. In the $\epsilon$-regime the
intrinsic statistical fluctuations of correlation functions turn out
to be particularly large. The signal can be greatly improved by making
specific use of the low modes of the Dirac operator
\citetwo{Giusti:2004yp}{Giusti:2004bf}. Results for the case $m_c=m_u$
will be published shortly.

\newabstract 
\begin{center}
{\large\bf Chiral Extrapolation of Baryon Properties---Recent Progress}\\[0.5cm]
Thomas R. Hemmert\\[0.3cm]
Physik Department T39, TU M{\" u}nchen, Germany \\[0.3cm]
\end{center}

The methods of chiral effective field theory (ChEFT) can be used to study
the quark-mass dependence of baryon properties. In this talk I have given an
 update regarding recent developments in this field. I have discussed
 predictions for the quark-mass dependence of the nucleon mass and the
 nucleon sigma term in relativistic/covariant Baryon Chiral Perturbation 
Theory (BChPT) to next-to-leading one-loop order (NLO) \cite{PHW}.
 Furthermore, I have presented evidence that such a calculation is
 in accordance with the principles of ChEFT out to relatively large
 pion masses around 500 MeV \cite{BHM}. An analysis
 of the uncertainty-/error-bands of this NLO calculation was presented
        \cite{Bernhard}.
 As a second topic I discussed the chiral extrapolation of the magnetic 
moments of the nucleon utilizing the methods of covariant/relativistic
 BChPT. It was argued that the previously observed breakdown of NLO
 calculations for pion-masses around 300 MeV is connected with an
 insufficient treatment of the quark-mass dependence of the analytic
 structures. A method of self-consistent propagators was presented to
 overcome this problem \cite{GH}. First results for the quark-mass
 dependence of the mass of Delta(1232) were also shown \cite{BHM2}. 
Finally, new predictions illuminating the role of explicit Delta(1232)
 degrees of freedom in the volume dependence of the mass of the
 nucleon \cite{Tim} and of the axial coupling of the nucleon \cite{QCDSF}
 were presented.

\newabstract 
\begin{center}
\vspace*{-2.0cm}
{\large\bf Progress, Challenges and Strategies in Lattice QCD${}^{\ast\ast}$ }
\\[0.5cm]
Elisabetta Pallante\\[0.3cm]
Centre for Theoretical Physics, Physics Dept.,
Groningen University,\\
Nijenborgh 4, 9747AG Groningen, The Netherlands.    \\
${}^{\ast\ast}$ {\em\small This talk can be found at 
http://www.itkp.uni-bonn.de/~eft04/talks/pallante.pdf}
\\[0.3cm]
\end{center}

Recent years have seen a substantial progress in Lattice QCD. 
After reaching a better comprehension of quenched diseases and related 
systematic uncertainties, we finally entered the {\em dynamical era}. 
A promising result is the determination of {\em real world} light and strange 
quark masses, $f_\pi, f_K, |V_{us}|$ by the MILC collaboration \cite{MILC}.
Progress continues with the optimization of four alternative 
formulations of lattice fermion actions: staggered, domain wall, overlap and 
twisted mass Wilson fermions \cite{dynamical}. Recently, staggered $\chi$PT 
and twisted mass $\chi$PT have been added to the list of 
frameworks to guide lattice extrapolations to the chiral and continuum limit
\cite{schpt}.
 
The challenge remains the determination of non-leptonic 
two-body weak decays, in particular $K\to\pi\pi$, where the
essential role of final state interactions (FSI) and their proper treatment in 
the computation of $\epsilon'/\epsilon$ was addressed in \cite{fsi}. 
Hence, the necessity of incorporating FSI into the lattice determination 
of $K\to\pi\pi$ led to lattice strategies at {\em finite volume} 
\cite{fv}, which overcome the Maiani-Testa theorem,  and for
different regimes of quark masses and volumes: the $p$-regime, 
the $\epsilon'$- and $\epsilon$-regime when approaching the chiral 
limit.

The charm quark and GIM mechanism are expected to play a central role in 
the $\Delta I=1/2$ rule. The enhancement should come from the {\em eye-like}
 Wick contraction, which is zero at $m_u=m_c$. Since Nature provided us with  
 $m_c>\Lambda_\chi$, we expect the bulk of the effect coming from that region.
That is why the study of an active charm  with mass 
$m_c <\Lambda_\chi$ (i.e. SU(4) $\chi$PT) \cite{activecharm}, while being an 
instructive exercise, will not explain the bulk of the physical effect, and 
as expected, will originate a numerically suppressed octet enhancement due to 
threshold effects $~m_c\log{m_c/\Lambda_\chi}$. The new generation of Teraflop 
computers might finally be enough to attack the $\Delta I=1/2$ problem from 
first principles.

\newabstract 

\begin{center}
{\large \textbf{Topics in Effective Field Theory}}

\textbf{Jiunn-Wei Chen}\\[0.3cm]
Department of Physics and National Center for Theoretical Sciences at Taipei,

National Taiwan University, Taipei, Taiwan\\[0.3cm]
\end{center}

Several recently effective field theory applications are discussed:

1) \textbf{Lattice theory for cold atom systems:} We construct a lattice
theory describing a system of interacting nonrelativistic spin s=1/2
fermions at nonzero chemical potential \cite{CK}. The theory is applicable
whenever the interparticle separation is large compared to the range of the
two-body potential, and does not suffer from a sign problem. In particular,
the theory could be useful in studying the thermodynamic limit of fermion
systems for which the scattering length is much larger than the
interparticle spacing, with applications to realistic atomic systems such as
the BEC to BCS transitions and dilute neutron gases.

2) \textbf{Inequalities for light nuclei in the Wigner symmetry limit:}
Using effective field theory we derive inequalities for light nuclei in the
Wigner (isospin and spin) symmetry limit \cite{CLS}. We prove that the
energy of any three-nucleon state is bounded below by the average energy of
the lowest two-nucleon and four-nucleon states. We show how this is modified
by lowest-order terms breaking Wigner symmetry and prove general energy
convexity results for SU(N). We also discuss the inclusion of
Wigner-symmetric three and four-nucleon force terms.

3) \textbf{Universality of the EMC effect:} Using effective field theory, we
investigate nuclear modification of nucleon parton distributions (for
example, the EMC effect) \cite{CD}. We show that the universality of the
shape distortion in nuclear parton distributions (the factorization of the
Bjorken $x$ and atomic number ($A$) dependence) is model independent and
emerges naturally in effective field theory. We then extend our analysis to
study the analogous nuclear modifications in isospin and spin dependent
parton distributions and generalized parton distributions.


\newabstract 
\begin{center}
{\large\bf Finite volume effects\\ using lattice chiral perturbation theory}\\[0.5cm]
{\bf B. Borasoy}$^1$ and R. Lewis$^2$\\[0.3cm]
$^1$Helmholtz-Institut f\"ur Strahlen und Kernphysik,\\
Universit\"at Bonn, Nu{\ss}allee 14-16, D-53115 Bonn, Germany\\[0.3cm]
$^2$Department of Physics,\\
University of Regina, Regina, SK, S4S 0A2,Canada\\[0.3cm]
\end{center}

The physics of pions within a finite volume is explored using
lattice regularized chiral perturbation theory. 
This regularization scheme permits a straightforward computational
approach to be used in place of analytical continuum techniques.

The continuum limit must be identical to
any viable continuum regulator, but lattice regularization has the feature
of being easy to manage numerically.
A suggested advantage of this regularization scheme is that
the renormalization can be carried out numerically, leaving fewer analytical
steps to be performed. 
Beginning from a Lagrangian that displays the lattice spacing explicitly
and also maintains exact chiral symmetry \citetwo{LO}{BLO}
one can simply derive the Feynman
propagators and vertices then type those directly into a computer program.
Loop diagrams are just summations of a finite number of momentum values and
the numerics are finite at every step.
For a sufficiently small lattice spacing, observables must be independent of
the lattice spacing.

Using the pion mass, decay constant, form factor and charge radius as examples,
it is shown how numerical results for volume dependences are obtained at the
one-loop level \citetwo{BLM}{BL}.
The expressions for the pion mass and decay constant are known in dimensional regularization
\cite{GLvolume1}, and results from the two regularization schemes agree numerically.

\newabstract 
\begin{center}
{\large\bf Finite Volume effects for Decay Constants}\\[0.5cm]
Christoph Haefeli$^1$\\[0.3cm]
$^1$Institut f\"ur Theoretische Physik, Universit\"at Bern\\
\end{center}
This talk is based on joint work with G.~Colangelo and
S.~D\"urr \cite{Colangelo:2004xr}.

Lattice calculations are performed in a finite volume 
and since the extrapolation to the infinite volume is numerically quite
expensive, it is deserving that one may rely on analytical methods. In the
$p$-regime, i.e. $M_\pi L \gg 1$, the
asymptotic formula of L\"uscher \cite{Luscher:1985dn} offers an efficient way
to investigate these effects. It has been applied within the framework of CHPT
to the pion mass \cite{Colangelo:2003hf} and has been extended also to decay
constants \cite{Colangelo:2004xr}. For the pion decay constant, it reads
\begin{equation}
  F_\pi(L)-F_\pi = \frac{3}{8 \pi^2 M_\pi L} \int_{-\infty}^{\infty}d y
\; e^{-\sqrt{M_\pi^2+y^2} L} N_F(iy) +O(e^{-\bar ML}) \, ,
\label{eq:Fpi}
\end{equation}
where $F_\pi(L)$ is the pion decay constant in finite volume, $\bar M
\geq \sqrt{3/2}M_\pi$ and $N_F(\nu)$ denotes the subtracted infinite volume
forward scattering amplitude of the matrix element with three pions created
out of the vacuum with an axial current. By employing
the chiral expansion of $N_F(\nu)$ up to NLO, we were able to investigate
Eq.(\ref{eq:Fpi}) numerically. The main results are:
\begin{itemize}
\item the finite volume corrections are exponentially suppressed for  large
  values of $M_\pi L$ and become negligible rather quickly;
\item the leading term in the chiral expansion of the asymptotic formula
  receives large corrections even for the physical values of the quark masses;
\item the contributions from the cut terms in $N_F$ turn out to be 
  very small, such that even a second order polynomial for the amplitude $N_F$
  is already a good approximation. The finite volume shift can then be
  written in a very compact manner, in terms of two modified
  Bessel functions. 
\end{itemize}
I have also discussed an extension of Eq.(\ref{eq:Fpi}), which allows to
estimate sub leading contributions.

\newabstract 
\begin{center}
{\large\bf Isospin violation in semileptonic decays}\\[0.5cm] 
{\bf Helmut Neufeld}\\[0.3cm]
Institut f\"ur Theoretische Physik der Universit\"at Wien,\\
Boltzmanngasse 5, A-1090 Wien, Austria.\\[0.3cm]
\end{center}

The appropriate theoretical framework for the treatment of 
isospin-violating effects in semileptonic decays
is provided by chiral perturbation theory with virtual photons and leptons 
\cite{KNRT00}. This effective quantum field theory 
describes the interactions of the pseudoscalar octet, the photon and the 
light leptons at low energies and allows a comparison of experimental data 
with the predictions of the standard model. It has been applied for the 
analysis of semileptonic kaon \cite{Kl3} and pion \cite{pibeta} decays.

The combined analysis of $K^0_{e 3}$ and $K^+_{e 3}$ data 
may serve as an illustration. The standard model allows for a remarkably 
precise prediction of the quantity
$r_{+0} := f_+^{K^+ \pi^0}(0) / f_+^{K^0 \pi^-}(0)$.
This quantity is  largely insensitive to the dominating theoretical 
uncertainties, in particular the contributions of order $p^6$. The 
theoretical prediction \cite{CKM}, $r_{+0}^{\rm th} = 1.022 \pm 0.003 -16 
\pi \alpha X_1$,
depends only on the (unknown) electromagnetic low-energy coupling $X_1$ 
\cite{KNRT00}. 
Already simple dimensional analysis ($|X_1| \le 1/(4 \pi)^2$) confines 
$r_{+ 0}$ to the rather narrow band 
$1.017  \leq r_{+0}^{\rm th} \leq 1.027$, 
leading to a  stringent test for the corresponding observable quantity
$r_{+ 0}^{\exp}$ \cite{CKM}. 
Using the most recent $K^+$ and $K_L$ data \cite{Litov}, one finds 
$r_{+ 0}^{\exp} = 1.038 \pm 0.007 \, (1.036 \pm 0.008)$
where linear (quadratic) form factor fits have been used. 
The corresponding numbers using $K_S$ data read
$r_{+ 0}^{\exp} = 1.036 \pm 0.010 \, (1.035 \pm 0.011)$.

The (small) discrepancy between present data and the standard model 
prediction  should encourage further experimental and 
theoretical efforts on this issue.


\begin{thebibliography}{12}

\bibitem{JM1}
E.~Jenkins and A.~V.~Manohar,
Phys.\ Rev.\ Lett.\  {\bf 93}, 022001 (2004)
[arXiv:hep-ph/0401190].

\bibitem{JM2}
E.~Jenkins and A.~V.~Manohar,
JHEP {\bf 0406}, 039 (2004)
[arXiv:hep-ph/0402024].

\bibitem{JM3}
E.~Jenkins and A.~V.~Manohar,
Phys.\ Rev.\ D {\bf 70}, 034023 (2004)
[arXiv:hep-ph/0402150].

\bibitem{AM1}
A.~V.~Manohar,
Phys.\ Rev.\ D {\bf 70} (2004) 014004
[arXiv:hep-ph/0404122].

\bibitem{AM2}
A.~V.~Manohar,
Nucl.\ Phys.\ B {\bf 248} (1984) 19.

\end{thebibliography}

\begin{thebibliography}{12}
\bibitem{Mexico}H.~Leutwyler,
{\it $\pi \pi$ scattering: theory is ahead of experiment}, talk given at the
{\it 10$th$ Mexican School of Particles and Fields}, 
arXiv:hep-ph/0212323.

\bibitem{Colangelo Villasimius}
G.~Colangelo,
{\it $\pi \pi$ scattering, pion form factors and chiral perturbation theory},
arXiv:hep-ph/0501107.

\bibitem{ACCGL}B.~Ananthanarayan, I.~Caprini, G.~Colangelo, J.~Gasser and
H.~Leutwyler, 
{\it Scalar form factors of light mesons},
arXiv:hep-ph/0409222.

\bibitem{CCGL} I.~Caprini, G.~Colangelo, J.~Gasser and
H.~Leutwyler, {\it On the precision of\\ the theoretical
  predictions for $\pi\pi$ scattering}, arXiv:hep-ph/0306122;\\ H.~Leutwyler,
  http://benasque.ecm.ub.es/2004quarks/2004quarks.html
\end{thebibliography}

\begin{thebibliography}{9}

\bibitem{param}
S.~Descotes-Genon, L.~Girlanda and J.~Stern,
JHEP {\bf 0001}, 041 (2000).

\bibitem{resum}
S.~Descotes-Genon \emph{et al.},
Eur.\ Phys.\ J.\ C {\bf 34}, 201 (2004).

\bibitem{uuss}
B.~Moussallam,
Eur.\ Phys.\ J.\ C {\bf 14} (2000) 111;
JHEP {\bf 0008} (2000) 005.\\
S.~Descotes-Genon,
JHEP {\bf 0103} (2001) 002.\\
P.~B\"uttiker \emph{et al.},
Eur.\ Phys.\ J.\ C {\bf 33} (2004) 409.


\bibitem{extrapol}
S.~Descotes-Genon,
hep-ph/0410233, to be published in Eur.\ Phys.\ J.\ C.


\bibitem{finvol}
D.~Becirevic and G.~Villadoro,
Phys.\ Rev.\ D {\bf 69}, 054010 (2004).\\
G.~Colangelo and S.~D\"urr,
Eur.\ Phys.\ J.\ C {\bf 33}, 543 (2004).\\
G.~Colangelo and C.~Haefeli,
Phys.\ Lett.\ B {\bf 590}, 258 (2004).

\end{thebibliography}

\begin{thebibliography}{12}

\bibitem{gdh} S. B. Gerasimov, Sov. J. Nucl. Phys. {\bf 2}, 430
(1966); D.D. Drell, A.C. HJearn, Phys. Rev. Lett. {\bf 16}, 908
(1966).
\bibitem{acm} G. Altarelli, N. Caabibbo, nd L. Maiani, Phys. Lett.
{\bf B40}, 415 (1972).
\bibitem{gdh1} V. Pascalutsa, B.R. Holstein, and M. Vanderhaeghen,
Phys. Lett. {\bf B600}, 239 (2004).
\end{thebibliography}

\begin{thebibliography}{12}

\bibitem{PSI}
D.~Chatellard {\it et al.},
Nucl.\ Phys.\ A {\bf 625} (1997) 855;
P.~Hauser {\it et al.},
Phys.\ Rev.\ C {\bf 58} (1998) 1869.

\bibitem{Meissner}
S.~R.~Beane, V.~Bernard, T.-S.~H.~Lee and U.-G.~Mei{\ss}ner,
Phys.\ Rev.\ C {\bf 57} (1998) 424
[arXiv:nucl-th/9708035].
S.~R.~Beane, V.~Bernard, E.~Epelbaum, U.-G.~Mei{\ss}ner and D.~R.~Phillips,
Nucl.\ Phys.\ A {\bf 720} (2003) 399
[arXiv:hep-ph/0206219].



\bibitem{Borasoy}
B.~Borasoy and H.~W.~Grie{\ss}hammer,
Int.\ J.\ Mod.\ Phys.\ E {\bf 12} (2003) 65 [arXiv:nucl-th/0105048].



\bibitem{Beane}
S.~R.~Beane and M.~J.~Savage,
Nucl.\ Phys.\ A {\bf 717} (2003) 104
[arXiv:nucl-th/0204046].

\bibitem{deuteron}
U.-G. Mei{\ss}ner, U. Raha and A. Rusetsky, in preparation


\end{thebibliography}

\begin{thebibliography}{99}

\bibitem{Baldini} 
R. Baldini et al. [DEAR Collaboration], DAPHNE exotic atom research: The DEAR
proposal, LNF-95-055-IR.

\bibitem{Cargnelli}
M. Cargnelli et al. [DEAR Collaboration], in Proceedings of the HadAtom03
Workshop, 12-17 October 2003,ECT(Trento, Italy),
[arXiv:hep-ph/0401204].

\bibitem{Gall}
A. Gall, J. Gasser, V. E. Lyubovitskij, A. Rusetsky,
Phys. Lett. B {\bf 462} (1999) 335
[arXiv:hep-ph/9905309].

\bibitem{Lyubovitskij}
V. E. Lyubovitskij, A. Rusetsky,
Phys. Lett. B {\bf 494} (2000) 9
[arXiv:hep-ph/0009206].

\bibitem{Zemp}
P. Zemp, in Proceedings of HadAtom03 Workshop, 13-17 October 2003 ECT(Trento
Italy), J. Gasser, P. Zemp, in preparation,
[arXiv:hep-ph/0009206].

\bibitem{Schweizer}
J. Schweizer,
Eur. Phys. J. C {\bf 36} (2004) 483
[arXiv:hep-ph/0405034].

\bibitem{Meissner}
U.-G. Mei\ss ner, U. Raha, and A. Rusetsky
Eur. Phys. J. C {\bf 35} (2004) 249-357
[arXiv:hep-ph/0402261].

\end{thebibliography}

\begin{thebibliography}{12}
\bibitem{BN1} B.~Borasoy and R.~Ni{\ss}ler, Eur.\ Phys.\ J.\ {\bf A19} (2004) 367
\bibitem{BN2} B.~Borasoy and R.~Ni{\ss}ler, Nucl.\ Phys.\ {\bf A740} (2004) 362
\end{thebibliography}

\begin{thebibliography}{12}

\bibitem{GKPV} 
J.~Gasser, B.~Kubis, N.~Paver and M.~Verbeni,
arXiv:hep-ph/0412130.

\bibitem{FFS} 
H.~W.~Fearing, E.~Fischbach and J.~Smith,
Phys.\ Rev.\ D {\bf 2} (1970) 542.

\bibitem{BEG} 
J.~Bijnens, G.~Ecker and J.~Gasser,
Nucl.\ Phys.\ B {\bf 396} (1993) 81.

\bibitem{na48}
A.~Lai {\it et al.}  [NA48 Collaboration],
Phys.\ Lett.\ B {\bf 605} (2005) 247.

\bibitem{ktev} 
A.~Alavi-Harati {\it et al.}  [KTeV Collaboration],
Phys.\ Rev.\ D {\bf 64} (2001) 112004.

\bibitem{ktev_04}
T.~Alexopoulos {\it et al.}  [KTeV Collaboration],
arXiv:hep-ex/0410070.
\end{thebibliography}

\begin{thebibliography}{12}
\bibitem{eglpr} 
G.~Ecker, J.~Gasser, A.~Pich and E.~de Rafael,
Nucl.\ Phys.\ B {\bf 321} (1989) 311,\ 
G.~Ecker, J.~Gasser, H.~Leutwyler, A.~Pich and E.~de Rafael,
Phys.\ Lett.\ B {\bf 223} (1989) 425.
\bibitem{mouss97} 
B.~Moussallam,
Nucl.\ Phys.\ B {\bf 504} (1997) 381
[arXiv:hep-ph/9701400].
\bibitem{grs}
J.~Gasser, A.~Rusetsky and I.~Scimemi,
Eur.\ Phys.\ J.\ C {\bf 32} (2003) 97
[arXiv:hep-ph/0305260].
\bibitem{am04}
B.~Ananthanarayan and B.~Moussallam,
JHEP {\bf 0406} (2004) 047
[arXiv:hep-ph/0405206].


\end{thebibliography}

\begin{thebibliography}{12}
\bibitem{prl1} E.M. Aitala et al., E791 Collaboration, Phys. Rev. Lett. 86
(2001) 765.
\bibitem{prl2}E.M. Aitala et al., E791 Collaboration, Phys. Rev. Lett. 86 (2001)
770.
\bibitem{prl3}E.M. Aitala et al., E791 Collaboration, Phys. Rev. Lett. 89 (2002)
127801.
\bibitem{meiss}S. Gardner and U.-G. Mei{\ss}ner, Phys. Rev. D65 (2002) 094004; 
J.A. Oller, eConf C0304052:WG412,2003; hep-ph/0306294.
\bibitem{npa}J.A. Oller and E. Oset, Nucl. Phys. {\bf A620} (1997) 438; (E) Nucl. Phys. {\bf A652} 
(1999) 407. 
\bibitem{jamin} M. Jamin, J.A. Oller and A. Pich, Nucl. Phys. {\bf B587} (2000) 331.
\end{thebibliography}

\begin{thebibliography}{99}


\bibitem{cola}
D.~Jido, J.A. Oller, E. Oset, A. Ramos, U.G. Mei{\ss}ner,
Nucl.\ Phys.\ A {\bf 725} (2003) 181.

\bibitem{nacher}
J.~C.~Nacher, E.~Oset, H.~Toki and A.~Ramos,
Phys.\ Lett.\ B {\bf 461} (1999) 299.

\bibitem{hyodo}
T.~Hyodo, A.~Hosaka, M.~J.~Vicente Vacas and E.~Oset,
Phys.\ Lett.\ B {\bf 593} (2004) 75.

\bibitem{sarkar}
S.~Sarkar, E.~Oset and M.~J.~V.~Vacas,
arXiv:nucl-th/0407025.

\bibitem{delka}
S.~Sarkar, E.~Oset and M.~J.~Vicente Vacas,
arXiv:nucl-th/0404023.

\end{thebibliography}

\begin{thebibliography}{12}
\bibitem{dalitz}
K.~Kampf, M.~Knecht and J.~Novotn\'y,
The Dalitz decay $\pi^0 \rightarrow e^+e^-\gamma$
revisited, in preparation.
\bibitem{Knecht:1999ag}
M.~Knecht, H.~Neufeld, H.~Rupertsberger and P.~Talavera,
Eur.\ Phys.\ J.\ C {\bf 12}, 469 (2000)
[arXiv:hep-ph/9909284].
\bibitem{Knecht:2001xc}
M.~Knecht and A.~Nyffeler,
Eur.\ Phys.\ J.\ C {\bf 21}, 659 (2001)
[arXiv:hep-ph/0106034].

\end{thebibliography}

\begin{thebibliography}{12}
\bibitem{DEHZ}
M.~Davier, S.~Eidelman, A.~Hocker and Z.~Zhang,
Eur.\ Phys.\ J.\ C {\bf 31} (2003) 503
[arXiv:hep-ph/0308213].

\bibitem{CGL} G.~Colangelo, J.~Gasser and H.~Leutwyler,
Nucl.\ Phys.\ B {\bf 603} (2001) 125
 [arXiv:hep-ph/0103088].

\bibitem{Eidelman:2003uh}
L.~Lukaszuk,
Phys.\ Lett.\ B {\bf 47} (1973) 51, and 
S.~Eidelman and L.~Lukaszuk,
Phys.\ Lett.\ B {\bf 582} (2004) 27
[arXiv:hep-ph/0311366].

\bibitem{Colangelo:2003yw}
G.~Colangelo,
arXiv:hep-ph/0312017.

\bibitem{Akhmetshin:2003zn}
R.~R.~Akhmetshin {\it et al.}  [CMD-2 Collaboration],
Phys.\ Lett.\ B {\bf 578} (2004) 285
[arXiv:hep-ex/0308008].

\bibitem{Aloisio:2004bu}
A.~Aloisio {\it et al.}  [KLOE Collaboration],
Phys.\ Lett.\ B {\bf 606} (2005) 12
[arXiv:hep-ex/0407048].

\end{thebibliography}

\begin{thebibliography}{12}

\bibitem{BG}
C.~W.~Bernard and M.~F.~L.~Golterman,
Phys.\ Rev.\ D {\bf 46}, 853 (1992), \\
Phys.\ Rev.\ D {\bf 49}, 486 (1994).

\bibitem{Sharpe}
S.~R.~Sharpe and N.~Shoresh,
Phys.\ Rev.\ D {\bf 62}, 094503 (2000), \\
Phys.\ Rev.\ D {\bf 64}, 114510 (2001).

\bibitem{BDL}
J.~Bijnens, N.~Danielsson and T.~A.~L\"ahde,
Phys.\ Rev.\ D {\bf 70}, 111503 (2004).

\bibitem{BL1}
J.~Bijnens and T.~A.~L\"ahde, hep-lat/0501014.

\bibitem{Latt}
C.~Aubin {\it et al.}  [MILC Collaboration],
Phys.\ Rev.\ D {\bf 70}, 114501 (2004).

\end{thebibliography}

\begin{thebibliography}{12}

\bibitem{PI:02} A. Pich, arXiv:hep-ph/0205030.

\bibitem{HO:74} G. 't Hooft, Nucl. Phys. B72 (1974) 461;
   B75 (1974) 461.

\bibitem{WI:79} E. Witten, Nucl. Phys. B160 (1979) 57.
  
\bibitem{EGPR:89} G. Ecker et al.,
 Nucl. Phys. B321 (1989) 311; 
 Phys. Lett. B223 (1989) 425.

\bibitem{MO:95} B. Moussallam, Phys. Rev. D51 (1995) 4939;
Nucl. Phys. B504 (1997) 381.

\bibitem{KPdR} M. Knecht and E. de Rafael, Phys. Lett. B424 (1998) 335;
S. Peris et al., 
 JHEP 05 (1998) 011; 01 (2002) 024;
 Phys. Rev. Lett. 86 (2001) 14.

\bibitem{KN:01} M. Knecht and A. Nyffeler, Eur. Phys. J. C21 (2001) 659.

\bibitem{RPP:03} P.D.~Ruiz-Femen\'{\i}a, A.~Pich and J.~Portol\'es,
 JHEP 07 (2003) 003.

\bibitem{CEEPP:04} V. Cirigliano et al., 
 Phys. Lett. B596 (2004) 96; work on progress.

\bibitem{BGLP:03} J. Bijnens, E. G\'amiz, E. Lipartia and J. Prades,
 JHEP 04 (2003) 055.

\bibitem{RSP:04} I. Rosell, J.J. Sanz-Cillero and A. Pich,
 JHEP 08 (2004) 042.

\end{thebibliography}

\begin{thebibliography}{12}
\bibitem{BL}
T.~Becher and H.~Leutwyler,
Eur.\ Phys.\ J.\ C {\bf 9} (1999) 643
[arXiv:hep-ph/9901384].\vs
\bibitem{BM} 
P.~C.~Bruns and U.-G.~Mei{\ss}ner,
arXiv:hep-ph/0411223, Eur. Phys. J. {\bf C} (2005) in print.\vs
\bibitem{CPPACS}
S.~Aoki {\it et al.}  [CP-PACS Collaboration],
Phys.\ Rev.\ D {\bf 60} (1999) 114508
[arXiv:hep-lat/9902018].\vs
\end{thebibliography}

\begin{thebibliography}{10}
\bibitem{BGLP} J. Bijnens, E. G\'amiz, E. Lipartia, and J. Prades,
J. High Energy Phys. 04 (2003) 055.
\bibitem{BGP1} J. Bijnens, E. G\'amiz, and J. Prades,
in preparation. 
\bibitem{BGP2} J. Bijnens, E. G\'amiz, and J. Prades,
in preparation. 
\end{thebibliography}

\begin{thebibliography}{12}
\bibitem{SFM} M.~Frink, U.-G.~Mei{\ss}ner and I.~Scheller,
{\em in preparation}.\vs
\bibitem{BHM}V.~Bernard, T.~R.~Hemmert and U.-G.~Mei{\ss}ner,
Nucl.\ Phys.\ A {\bf 732} (2004) 149
[arXiv:hep-ph/0307115].\vs
\bibitem{BM}B.~Borasoy and U.-G.~Mei{\ss}ner,
Annals Phys.\  {\bf 254} (1997) 192
[arXiv:hep-ph/9607432].\vs
\bibitem{FM}M.~Frink and U.-G.~Mei{\ss}ner,
JHEP {\bf 0407} (2004) 028
[arXiv:hep-lat/0404018].\vs

\bibitem{MILC}
C.~W.~Bernard {\it et al.},
Phys.\ Rev.\ D {\bf 64} (2001) 054506
[arXiv:hep-lat/0104002];
C.~Aubin {\it et al.},
Phys.\ Rev.\ D {\bf 70} (2004) 094505
[arXiv:hep-lat/0402030].\vs

\end{thebibliography}

\begin{thebibliography}{12}
\bibitem{vK1} 
U. van Kolck, Ph. D. dissertation, University of Texas (1993);
{\it Few--Body Syst. Suppl.} {\bf 9} (1995) 444.
\bibitem{vK2} 
J.L. Friar, U. van Kolck, M.C.M. Rentmeester, and R.G.E. Timmermans,
{\it Phys. Rev. C} {\bf 70} (2004) 044001. 
\bibitem{vK3}
U. van Kolck, J.L. Friar, and T. Goldman,
{\it Phys. Lett. B} {\bf 371} (1996) 169.
\bibitem{vK4}
J.L. Friar, U. van Kolck, G.L. Payne, and S.A. Coon,
{\it Phys. Rev. C}  {\bf 68} (2003) 024003;
J.A. Niskanen, {\it Phys. Rev. C}  {\bf 65} (2002) 037001.
\bibitem{vK7}
E. Epelbaum, U.-G.\ Mei{\ss}ner, and J.E.\ Palomar,
{\tt nucl-th/0407037}.
\bibitem{vK5}
J.L. Friar, G.L. Payne, and U. van Kolck, 
{\it Phys. Rev. C} (to appear), {\tt nucl-th/0408033}.
\bibitem{vK6}
U. van Kolck, J.A. Niskanen, and G.A. Miller,
{\it Phys.\ Lett. B} {\bf 493} (2000) 65.
\end{thebibliography}

\begin{thebibliography}{12}

\bibitem{Weinberg:rz}
S.~Weinberg,
Phys.\ Lett.\ B {\bf 251}, 288 (1990).

\bibitem{Gegelia:2004pz}
J.~Gegelia and S.~Scherer,
arXiv:nucl-th/0403052.

\bibitem{Djukanovic:2004px}
D.~Djukanovic, M.~R.~Schindler, J.~Gegelia and S.~Scherer,
arXiv:hep-ph/0407170.

\end{thebibliography}

\begin{thebibliography}{12}
\bibitem{ordonez}     C. Ord\'{o}\~{n}ez, L. Ray, U. van Kolck, Phys. Rev. C 53, 2086 (1996)  
\bibitem{entem} D.R. Entem, R. Machleidt, Phys. Rev. C 68,041001 (2003)
\bibitem{epelbaum}  E. Epelbaum, W. Gl\"ockle, U.-G. Mei{\ss}ner, Nucl. Phys. A 747, 362 (2005)
\bibitem{beane} S. R. Beane {\it et al.}, Nucl. Phys. A 700, 375 (2002)
\bibitem{arriola} M. Pav\'on Valderrama, and E. Ruiz Arriola,  Phys. Rev. C  70, 044006 (2004)
\bibitem{nogga} A. Nogga, R. Timmermanns, and U. van Kolck, in preparation
\bibitem{meissner} U.-G. Mei{\ss}ner,  INT Workshop on "Theories of Nuclear Forces and Few-Nucleon Systems", June 2001
\end{thebibliography}

\begin{thebibliography}{12}
\bibitem{PavonValderrama:2003np}
M.~Pavon Valderrama and E.~Ruiz Arriola,
Phys.\ Lett.\ B {\bf 580}, 149 (2004)

\bibitem{PavonValderrama:2004nb}
M.~Pavon Valderrama and E.~Ruiz Arriola,
Phys.\ Rev.\ C {\bf 70} (2004) 044006


\bibitem{PavonValderrama:2004se}
M.~Pavon Valderrama and E.~Ruiz Arriola,
arXiv:nucl-th/0407113.

\bibitem{PavonValderrama:2004td}
M.~Pavon Valderrama and E.~Ruiz Arriola,
arXiv:nucl-th/0410020.


\end{thebibliography}

\begin{thebibliography}{12}
\bibitem{rhoChPT}
J.~A.~Oller,
Phys.\ Rev.\ C {\bf 65}, 025204 (2002)
[arXiv:hep-ph/0101204].
U.~G.~Mei{\ss}ner, J.~A.~Oller and A.~Wirzba,
Annals Phys.\  {\bf 297}, 27 (2002)
[arXiv:nucl-th/0109026].

\bibitem{grwpaper}  L.~Girlanda, A.~Rusetsky and W.~Weise,
Annals Phys.\  {\bf 312} (2004) 92
[arXiv:hep-ph/0311128].

\end{thebibliography}

\begin{thebibliography}{12}
\bibitem{nucmat} N. Kaiser, S. Fritsch and W. Weise, \textit{Nucl. Phys.} 
\textbf{A697} (2002) 255.
\bibitem{deltamat} S. Fritsch, N. Kaiser and W. Weise, \textit{Nucl. Phys.} 
\textbf{A} (2005) in print; nucl-th/0406038.
\bibitem{pot} N. Kaiser, S. Fritsch and W. Weise, \textit{Nucl. Phys.} 
\textbf{A700} (2002) 343.
\bibitem{liquidgas} S. Fritsch, N. Kaiser and W. Weise, \textit{Phys. Lett.} 
\textbf{B545} (2002) 73.
\bibitem{spinstab} N. Kaiser, \textit{Phys. Rev.} \textbf{C70} (2004) 
054001. 
\end{thebibliography}

\begin{thebibliography}{12}

\bibitem{Wil71} 
K.G.~Wilson, Phys.\ Rev.\ D {\bf 3}, 1818 (1971).

\bibitem{BHK99} 
P.F.~Bedaque, H.-W.~Hammer, and U.~van Kolck,
Phys.\ Rev.\ Lett.\  {\bf 82}, 463 (1999).

\bibitem{BrH04}
E.~Braaten and H.-W.~Hammer,
arXiv:cond-mat/0410417.

\bibitem{BrH03} 
E.~Braaten and H.-W.~Hammer,
Phys.\ Rev.\ Lett.\  {\bf 91}, 102002 (2003).

\bibitem{Hammer:2004as}
H.-W.~Hammer and D.~T.~Son,
Phys.\ Rev.\ Lett.\  {\bf 93}, 250408 (2004).

\end{thebibliography}

\begin{thebibliography}{12}
\bibitem{Platter:2004qn}
L.~Platter, H.~W.~Hammer and U.-G.~Mei\ss ner,
Phys. Rev. A {\bf 70}, 250 (2004).

\bibitem{Platter:2004zs}
L.~Platter, H.~W.~Hammer and U.-G.~Mei\ss ner,
arXiv:nucl-th/0409040, Phys. Lett. B, in print.

\bibitem{3boson}
P.F.~Bedaque, U.~van Kolck, and H.-W.~Hammer,
Nucl.\ Phys.\ A {\bf 646}, 444 (1999).


\end{thebibliography}

\begin{thebibliography}{12}
\bibitem{ref1} V.~Baru, C.~Hanhart, A.~E.~Kudryavtsev and U.~G.~Mei{\ss}ner,
Phys.\ Lett.\ B {\bf 589} (2004) 118.
\bibitem{inprep} V. Lensky et al., in preparation.
\end{thebibliography}

\begin{thebibliography}{12}
\bibitem{KBMOrtho}
H. Krebs, V. Bernard, Ulf-G. Mei\ss ner, arXiv: nucl-th/0407078,\\ 
accepted for publication in Annals of Physics
\bibitem{Suzuki1} K. Suzuki, R. Okamoto, Prog. Theor. Phys 70 (1983) 439
\bibitem{KBMq3}
H. Krebs, V. Bernard, Ulf-G. Mei\ss ner, Nucl. Phys. A 713 (2003) 405,\\
arXiv: nucl-th/0207072
\bibitem{Ewald}
I.~Ewald et al., Phys.\ Lett.\ B 499 (2001) 238,
arXiv: nucl-ex/0010008
\bibitem{KBMq4}
H. Krebs, V. Bernard, Ulf-G. Mei\ss ner, Eur. Phys. J. A 22 (2004) 503,\\ 
arXiv: nucl-th/0405006 
\bibitem{EpelbImpr2}
E. Epelbaum, W. Gl\"ockle, Ulf-G. Mei\ss ner, 
Eur. Phys. J. A 19 (2004) 401,\\ 
arXiv: nucl-th/0308010
\end{thebibliography}

\begin{thebibliography}{8}
\bibitem{Griesshammer:2001uw} H.~W.~Grie{\ss}hammer and T.~R.~Hemmert:
  \PRC\textbf{65} (2002), 045207 [nucl-th/0110006].
  
\bibitem{polas2} R.~P.~Hildebrandt, H.~W.~Grie{\ss}hammer, T.~R.~Hemmert and
  B.~Pasquini: \EPJA\textbf{20} (2004), 293
  [nucl-th/0307070].
  
\bibitem{beane} S.R.~Beane et al: \NPA\textbf{656} (1999), 367;
  \PLB\textbf{567} (2003), 200 and [nucl-th/0403088].
  
\bibitem{polasfromdeuteron} R.~P.~Hildebrandt, H.~W.~Grie{\ss}hammer,
  T.~R.~Hemmert and D.~R.~Phillips: 
  [nucl-th/0405077]. Accepted by \NPA.
  
\bibitem{polas3} R.~P.~Hildebrandt, H.~W.~Grie{\ss}hammer and T.~R.~Hemmert:
  \EPJA\textbf{20} (2004), 329 [nucl-th/0308054].
  
\end{thebibliography}

\begin{thebibliography}{12}
\bibitem{HS1}J. Hirn and J. Stern, Eur. Phys. J. {\bf C34}, 447, (2004);
[hep-ph/0401032]
\bibitem{HS2} J. Hirn and J. Stern, JHEP, {\bf 09}, 058, (2004);
[hep-ph/0403017].
\end{thebibliography}

\begin{thebibliography}{12}

\bibitem{Kunszt:2004ps}Z.~Kunszt, A.~Nyffeler and M.~Puchwein, JHEP {\bf 0403}, 061 (2004).

    \bibitem{Son:2003et}D.~T. Son and M.~A. Stephanov, Phys.\ Rev.\ D {\bf 69}, 065020 (2004).

  
  \bibitem{Hirn:2004ze}J. Hirn and J. Stern, Eur.\ Phys.\ J.\ C {\bf 34}, 447 (2004).
  
  \bibitem{Contino:2003ve}R. Contino, Y. Nomura, and A. Pomarol,
  Nucl.\ Phys.\ B {\bf 671}, 148 (2003).


  \bibitem{Barbieri:2003pr}R. Barbieri, A. Pomarol, and R. Rattazzi, Phys.\ Lett.\ B {\bf 591}, 141 (2004).


\bibitem{everybody}J. Hirn, A. Pich, N. Rius, A. Santamaria and V. Sanz, {\em work in progress}.

\end{thebibliography}

\begin{thebibliography}{12}
\bibitem{Nowak} M.A. Nowak, hep-ph/0407272
\bibitem{Barnes} T. Barnes, F.E. Close and H.J. Lipkin, Phys. Rev. {\bf D68}
  (2003) 054024.
\bibitem{Weinberg} S. Weinberg, Phys. Rev. {\bf 137} (1965) B672.
\bibitem{Baru} V. Baru et al., Phys. Lett. {\bf B586} (2004) 53.
 \end{thebibliography}

\begin{thebibliography}{12}

\bibitem{Beane:2003da}
S.~R.~Beane, P.~F.~Bedaque, A.~Parreno and M.~J.~Savage,
Phys.\ Lett.\ B {\bf 585}, 106 (2004)
[arXiv:hep-lat/0312004].

\bibitem{Detmold:2004qn}
W.~Detmold and M.~J.~Savage,
Nucl.\ Phys.\ A {\bf 743}, 170 (2004)
[arXiv:hep-lat/0403005].

\bibitem{Beane:2003yx}
S.~R.~Beane, P.~F.~Bedaque, A.~Parreno and M.~J.~Savage,
Nucl.\ Phys.\ A {\bf 747}, 55 (2005)
[arXiv:nucl-th/0311027].


\end{thebibliography}

\begin{thebibliography}{12}

\bibitem{Maiani:ca}
L.~Maiani and M.~Testa,
{\it Phys. Lett.} {\bf B245}, 585 (1990).

\bibitem{Luscher:1990ux}
M.~L{\"u}scher,
{\it Nucl. Phys.} {\bf B354}, 531 (1991).

\bibitem{Fukugita:1994ve}
M.~Fukugita {\it et al.}, 
{\it Phys. Rev.} {\bf D52}, 3003 (1995).

\bibitem{Beane:2003da}
S.R.~Beane {\it et al.},
{\it Phys. Lett.} {\bf B585}, 106 (2004).

\bibitem{nplqcd}
 http://www-nsdth.lbl.gov/$\sim$bedaque/nplqcd

\end{thebibliography}

\vspace{-4mm}

\begin{thebibliography}{5}

\vspace{-3mm}
\itemsep -3pt

\bibitem{Gasser:1983yg}
J.~Gasser and H.~Leutwyler,
Annals Phys.\  {\bf 158} (1984) 142.

\bibitem{Durr:2004ta}
S.~D\"urr and C.~Hoelbling,
hep-lat/0411022.

\bibitem{AliKhan:2001tx}
A.~Ali Khan {\it et al.}  [CP-PACS Collab.],
Phys.\ Rev.\ D {\bf 65} (2002) 054505
[Erratum-ibid.\ D {\bf 67} (2003) 059901]
[hep-lat/0105015].

\bibitem{Durr}
S.~D\"urr,
Eur.\ Phys.\ J.\ C {\bf 29} (2003) 383 [hep-lat/0208051],
[hep-ph/0209319].

\bibitem{Aoki:2003yv}
S.~Aoki,
Phys.\ Rev.\ D {\bf 68} (2003) 054508 [hep-lat/0306027].

\end{thebibliography}

\begin{thebibliography}{12}
\bibitem{ref1}
W.~Bietenholz, S.~Capitani, T.~Chiarappa, N.~Christian, M.~Hasenbusch, K.~Jansen, K.-I.~Nagai, M.~Papinutto, L.~Scorzato, S.~Shcheredin, A.~Shindler, C.~Urbach, U.~Wenger and I.~Wetzorke,
JHEP 0412 (2004) 044.
\bibitem{ref2}
T.~Chiarappa, K.~Jansen, K.-I.~Nagai, M.~Papinutto, L.~Scorzato, A.~Shindler, C.~Urbach, U.~Wenger and I.~Wetzorke, hep-lat/0409107.
\bibitem{ref3}
F.~Farchioni, K.~Jansen, I.~Montvay, E.~Scholz, L.~Scorzato, A.~Shindler, N.~Ukita, C.~Urbach and I.~Wetzorke
hep-lat/0410031, to be published in EPJC.
\bibitem{ref4}
F.~Farchioni, R.~Frezzotti, K.~Jansen, I.~Montvay, G.C.~Rossi, E.~Scholz, A.~Shindler, N.~Ukita, C.~Urbach and I.~Wetzorke, hep-lat/0406039 to be published in
EPJC. 
\bibitem{ref5}
M. Guagnelli, K. Jansen, F. Palombi, R. Petronzio, A. Shindler and I. Wetzorke,
Phys.Lett. B597 (2004) 216.
\bibitem{ref6}
I. Wetzorke, K. Jansen, F. Palombi and A. Shindler, hep-lat/0409142.
\end{thebibliography}

\begin{thebibliography}{12}

\bibitem{mgys} M.~Golterman and Y.~Shamir,
arXiv:hep-lat/0411007.
\bibitem{rbcdyn} Y.~Aoki {\it et al.},
arXiv:hep-lat/0411006.
\bibitem{rbc} For quenched domain-wall fermions, see
T.~Blum {\it et al.}  [RBC Collaboration],
Phys.\ Rev.\ D {\bf 68}, 114506 (2003)
[arXiv:hep-lat/0110075];
J.~I.~Noaki {\it et al.}  [CP-PACS Collaboration],
Phys.\ Rev.\ D {\bf 68}, 014501 (2003)
[arXiv:hep-lat/0108013].

\end{thebibliography}

\begin{thebibliography}{12}


\bibitem{chirimp} 
C.~Gattringer, Phys.~Rev. D~{\bf 63} (2001) 114501;
C.\ Gattringer, I.\ Hip, C.\ B.\ Lang, Nucl.\ Phys.\ B~{\bf 597} (2001) 451.
\bibitem{broemmeletal} D. Br\"ommel {\it et al},
Phys.\ Rev.\ D~{\bf 69} (2004) 094513;
Nucl.\ Phys.\ B Proc.\ Suppl.\  {\bf 129-130} (2004) 251.
\bibitem{prd-paper}
T. Burch {\it et al.}, Phys.\  Rev.\  D {\bf 70} (2004)  054502;
see also hep-lat/0409014 and nucl-th-0501025.
\end{thebibliography}

\begin{thebibliography}{12}

\bibitem{Giusti:2004an}
L.\,Giusti, P.\,Hern\'andez, M.\,Laine, P.\,Weisz and H.\,Wittig,
JHEP {\bf 0411} (2004) 016;
P.\,Hern\'andez and M.\,Laine,
JHEP {\bf 0409} (2004) 018

\bibitem{Giusti:2002sm}
L.\,Giusti, C.\,Hoelbling, M.\,L\"uscher and H.\,Wittig,
Comput.\ Phys.\ Commun.\  {\bf 153} (2003) 31

\bibitem{Giusti:2004yp}
L.\,Giusti, P.\,Hern\'andez, M.\,Laine, P.\,Weisz and H.\,Wittig,
JHEP {\bf 0404} (2004) 013

\bibitem{Giusti:2004bf}
L.\,Giusti, P.\,Hern\'andez, M.\,Laine, C.\,Pena, P.\,Weisz,
J.\,Wennekers and H.\,Wittig,
hep-lat/0409031

\end{thebibliography}

\begin{thebibliography}{12}
\bibitem{PHW} M. Procura, T.R. Hemmert and W. Weise, Phys. Rev. D69, 034505 (2004).
\bibitem{BHM} V. Bernard, T.R. Hemmert and U.-G. Mei{\ss}ner, Nucl. Phys. A732, 149 (2004).
\bibitem{Bernhard} B. Musch, Diploma Thesis, TU M{\" u}nchen, in progress.
\bibitem{GH} T. Gail and T.R. Hemmert, forthcoming.
\bibitem{BHM2} V. Bernard, T.R. Hemmert and U.-G. Mei{\ss}ner, forthcoming.
\bibitem{Tim} T. Wollenweber, Diploma Thesis, TU M{\" u}nchen, January 2005.
\bibitem{QCDSF} QCDSF collaboration, in preparation. 
\end{thebibliography}

\begin{thebibliography}{12}
\bibitem{MILC}C. Aubin et al., Phys. Rev. D70 (2004)114501
\bibitem{dynamical}R. Frezzotti, hep-lat/0409138;
W. Bietenholz et al., hep-lat/0411001
\bibitem{schpt}C. Bernard, hep-lat/0412030; 
S. Sharpe, J.M.S. Wu, hep-lat/0411021
\bibitem{fsi}E. Pallante, A. Pich, I. Scimemi, Nucl. Phys. B617 (2001) 441
\bibitem{fv}L. Lellouch, M. L\"uscher, Comm. Math. Phys. 219 (2001) 31; 
C.-J.D. Lin et al., Phys. Lett. B581 (2004) 207
\bibitem{activecharm}L. Giusti et al., JHEP 0411 (2004) 016

\end{thebibliography}

\begin{thebibliography}{9}
\bibitem{CK} J.W. Chen and D.B. Kaplan, Phys. Rev. Lett. \textbf{92}, 257002
(2004).

\bibitem{CLS} J.W. Chen, D. Lee and T. Schaefer, Phys. Rev. Lett. \textbf{93}%
, 242302 (2004).

\bibitem{CD} J.W. Chen and W. Detmold, hep-ph/0412119.
\end{thebibliography}

\begin{thebibliography}{12}
\bibitem{LO} R.\ Lewis and P.\ P.\ A.\ Ouimet, Phys.\ Rev.\ D {\bf 64}, 034005
             (2001).
\bibitem{BLO} B.\ Borasoy, R.\ Lewis and P.\ P.\ A.\ Ouimet, 
              Phys.\ Rev.\ D {\bf 65}, 114023 (2002);
              B.\ Borasoy, R.\ Lewis and P.\ P.\ A.\ Ouimet, 
              Nucl.\ Phys.\ (Proc.Suppl.) {\bf 128}, 141 (2004).
\bibitem{BLM} B.\ Borasoy, R.\ Lewis and D.\ Mazur, hep-lat/0408040.
\bibitem{BL} B.\ Borasoy and R.\ Lewis, hep-lat/0410042.
\bibitem{GLvolume1} J.\ Gasser and H.\ Leutwyler, Phys.\ Lett.\ B {\bf 184},
                    83 (1987);\\
                    G.\ Colangelo and S.\ D\"urr, Eur.\ Phys.\ J.\ C {\bf 33},
                 543 (2004).
\end{thebibliography}

\begin{thebibliography}{12}
\bibitem{Colangelo:2004xr}
G.~Colangelo and C.~Haefeli,
Phys.\ Lett.\ B {\bf 590} (2004) 258
[arXiv:hep-lat/0403025].\\
G.~Colangelo, S.~D\"urr and C.~Haefeli, work in progress.

\bibitem{Luscher:1985dn}
M.~Luscher,
Commun.\ Math.\ Phys.\  {\bf 104} (1986) 177.

\bibitem{Colangelo:2003hf}
G.~Colangelo and S.~Durr,
Eur.\ Phys.\ J.\ C {\bf 33} (2004) 543
[arXiv:hep-lat/0311023].

\end{thebibliography}

\begin{thebibliography}{12}

\bibitem{KNRT00} 
M. Knecht, H. Neufeld, H. Rupertsberger and P. Talavera,
Eur. Phys. J. C 12 (2000) 469.

\bibitem{Kl3}
V. Cirigliano, M. Knecht, H. Neufeld, H. Rupertsberger and P. Talavera,
Eur. Phys. J. C 23 (2002) 121.

\bibitem{pibeta}
V. Cirigliano, M. Knecht, H. Neufeld and H. Pichl, Eur. Phys. J. C 27 
(2003) 255.

\bibitem{CKM}
V. Cirigliano, H. Neufeld and H. Pichl, Eur. Phys. J. C 35 (2004) 53.

\bibitem{Litov}
L. Litov, private communication.

\end{thebibliography}
\end{document}